\documentclass[%
reprint,
nofootinbib,
amsmath,
amssymb,
prd,
longbibliography,
superscriptaddress,twocolumn,superscriptaddress,floats,floatfix,showpacs]{revtex4-2}

\usepackage{booktabs}
\usepackage{verbatim}
\usepackage{mathrsfs}
\usepackage{graphicx}
\usepackage{dcolumn}
\usepackage{bm}
\usepackage{lipsum}
\usepackage{graphics}
\usepackage{bm}
\usepackage{bbm}
\usepackage{color}
\usepackage{microtype}
\usepackage{IEEEtrantools}
\usepackage{mathrsfs}
\usepackage{amsthm}
\usepackage{enumitem}
\usepackage[normalem]{ulem}
\usepackage{graphics}
\usepackage[T1]{fontenc}

\usepackage[dvipsnames, usenames]{xcolor}
\usepackage[unicode, colorlinks=true, linkcolor=linkcolor, citecolor=linkcolor, filecolor=linkcolor,urlcolor=linkcolor, pdfusetitle]{hyperref}
\usepackage[capitalise]{cleveref}
\usepackage[all]{hypcap}
\usepackage[utf8]{inputenc}
\usepackage{orcidlink}

\definecolor{linkcolor}{rgb}{0.0,0.3,0.5}
\definecolor{dodgerblue}{HTML}{1E90FF}
\definecolor{darkgreen}{rgb}{0,0.5,0}
\definecolor{romared}{RGB}{142,0,28}

\usepackage{hyperref}
\hypersetup{
    pdfnewwindow=true,     
    colorlinks=true,     
    linkcolor=romared,          
    citecolor=romared,        
    filecolor=romared,      
    urlcolor=romared        
}

\newcommand{\GSSI}{Gran Sasso Science Institute (GSSI), I-67100 L’Aquila, Italy}
\newcommand{\GranSasso}{INFN, Laboratori Nazionali del Gran Sasso, I-67100 Assergi, Italy}

\newcommand{\dd}{\mathrm{d}}
\newcommand{\tn}{\textnormal}

\usepackage[T1]{fontenc}

\begin{document}

\title{The significance of first post-adiabatic contributions for scalar charge measurements with intermediate and extreme mass ratio inspirals}

\author{Susanna Barsanti\,\orcidlink{0000-0001-7321-2512}}
\email{susanna.barsanti@ucd.ie}
\affiliation{School of Mathematics and Statistics, University College Dublin, Belfield, Dublin 4, Ireland}
\author{Ollie Burke\,\orcidlink{0000-0003-2393-209X}}
\affiliation{School of Physics and Astronomy, University of Glasgow, Glasgow G12 8QQ, UK}
\author{Andrea Maselli \orcidlink{0000-0001-8515-8525}}
\affiliation{\GSSI}
\affiliation{\GranSasso}
\author{Thomas P. Sotiriou \orcidlink{0000-0002-9089-4866}}
\affiliation{Nottingham Centre of Gravity \& School of Mathematical Sciences, University of Nottingham, University Park, Nottingham, NG7 2RD, UK}
\affiliation{School of Physics and Astronomy, University of Nottingham, University Park, Nottingham, NG7 2RD, UK}
\author{Andrew Spiers \orcidlink{https://orcid.org/0000-0003-0222-7578}}
\affiliation{School of Mathematics and Statistics, University College Dublin, Belfield, Dublin 4, Ireland}
\author{Niels Warburton \orcidlink{https://orcid.org/0000-0003-0914-8645}}
\affiliation{School of Mathematics and Statistics, University College Dublin, Belfield, Dublin 4, Ireland}

\begin{abstract}
 We present the first self-force-based beyond-GR waveform model incorporating post-adiabatic orbital evolution for intermediate- and extreme-mass-ratio inspirals in theories of gravity with additional scalar fields. Focusing on quasi-circular inspirals into a non-spinning primary, we combine a first post-adiabatic (1PA) gravitational sector with leading-order scalar field effects and use Bayesian injection–recovery studies to assess the impact of waveform systematics on the inference of scalar charges with LISA.
We find that neglecting 1PA effects in 
the gravitational sector can bias the inference 
of intrinsic binary parameters, while scalar-charge measurements remain robust across a 
wide range of mass ratios. 
In contrast, analysing signals from binaries in which the secondary carries a scalar charge using pure-GR templates leads 
to significant biases and underestimated 
uncertainties due to unmodelled correlations 
between the scalar charge and the binary parameters. 
We also investigate the role of 
secondary spin and find no significant correlation between the secondary 
spin and the scalar charge. Notably, up to 
a mass ratio of $10^{-4}$, the secondary spin 
itself remains unconstrained even in the pure-GR 
case, in contrast with previous claims in the 
literature. Finally, we show that modelling scalar emission with a leading-order dipolar post-Newtonian approximation -- for quasi-circular inspirals into a non-spinning primary -- introduces negligible systematic errors relative to fully relativistic scalar fluxes. 
\end{abstract}

\maketitle

\section{\label{sec:level1} Introduction}

The Laser Interferometer Space Antenna (LISA) will 
open a new window onto the low-frequency, millihertz 
gravitational-wave (GW) spectrum, enabling observations 
of previously unexplored classes of compact binaries, 
including asymmetric systems such as extreme- and 
intermediate-mass-ratio inspirals (EMRIs and IMRIs). 
These long-lived signals, produced by a compact 
object (the secondary) spiralling into a massive 
black hole (BH) (the primary), can accumulate up to 
$\mathcal{O}(10^5)$ GW cycles over several years 
of observation. This large number of cycles makes 
these systems exceptionally sensitive probes of 
fundamental physics in the strong-gravity regime~\cite{Cardenas-Avendano:2024mqp,Barack:2018yly,Barausse:2020rsu,Barausse:2016eii,Blazquez-Salcedo:2016enn,Glampedakis:2005cf,Barack:2006pq,Cardoso:2018zhm,Datta:2019epe,Pani:2019cyc,Maggio:2021uge,Destounis:2020kss,Piovano:2020ooe,Sago:2021iku,Piovano:2022ojl,Collodel:2021jwi}. The 
expected precision in parameter estimation~\cite{Babak:2017tow,Speri:2026ade} 
will, in particular, allow stringent tests of 
general relativity (GR).

New fundamental fields coupled for gravity could endow compact objects with new charges and affect their structure and dynamics. This could, in turn, leave characteristic imprints on the GW signal from binary coalescences. 
A prominent example is dipolar radiation sourced by the difference between the charges of the compact objects, which is absent in 
GR. The additional loss of energy into this new polarisation enhances the inspiral rate and alters the frequency and phase evolution of the conventional polarisations. Current constraints on such deviations largely 
come from the LIGO/Virgo/KAGRA Collaboration, which 
has used observations of comparable-mass binaries 
to bound departures from GR, including scenarios 
involving charged black holes. These analyses typically search for deviations in the post-Newtonian 
(PN) expansion of the inspiral phase, where leading 
beyond-GR effects often enter at $-1$PN order through 
dipolar emission~\cite{LIGOScientific:2026qni,LIGOScientific:2026fcf,LIGOScientific:2025wao,Sanger:2024axs}.

However, this approach is intrinsically limited by 
the need to extend PN descriptions to the  
strong-field regime, where PN models converge poorly. Numerical-relativity results 
suggest that nonlinear effects can partially mask or counterbalance deviations accumulated during the early inspiral~\cite{Corman:2025wun}, thereby complicating 
their interpretation.

In contrast, EMRIs and IMRIs provide a cleaner setting 
to probe such effects. Their dynamics can be 
described perturbatively in the mass ratio, and 
the self-force (SF) approach provides the most 
accurate framework for modelling asymmetric binaries\footnote{Recent second-order SF calculations in GR
suggest that the method can be highly accurate 
beyond its natural domain of EMRIs and IMRIs, 
and can provide a controlled framework to 
inform comparable-mass modelling~\cite{Wardell:2021fyy, Mathews:2025txc, Honet:2025dho, Kuchler:2024esj, Kuchler:2025hwx, Compere:2021iwh}, 
thereby improving tests of GR across different regimes. Additional hybridisation with results from PN can further the accuracy~\cite{Honet:2025lmk, Honet:2025gge}.}.

Exploiting this potential, however, requires 
unbiased inference from EMRI and IMRI observations. 
This in turn demands highly accurate waveform models 
and sophisticated data analysis techniques capable of 
exploring high-dimensional parameter spaces and handling 
exceptionally long-lived signals. In particular, avoiding systematic biases in parameter estimation requires waveform models that are 
accurate at the first post-adiabatic (1PA) order, which are constructed within a two-timescale description
of the orbital evolution~\cite{Miller:2020bft, Pound:2021qin} and incorporate second-order (in the mass ratio) 
dissipative contributions from the self-force. 

A comprehensive research effort is currently underway 
to produce 1PA waveforms suitable for LISA observations. 
First-order SF results in GR are now available 
for generic bound orbits, while second-order 
calculations remain limited to quasi-circular 
orbits around slowly spinning primaries~\cite{Mathews:2025txc}. 
These advances have enabled initial data analysis 
studies based on 1PA waveforms for quasi-circular 
inspirals into Schwarzschild black holes~\cite{Wardell:2021fyy}, 
with extensions including linear-in-spin corrections 
to the primary~\cite{Mathews:2025txc}. The importance of complete 
1PA models for LISA analyses have recently been quantified 
using Bayesian methods~\cite{Burke:2023lno}, which 
show that neglecting 1PA contributions can lead to 
significant systematic biases in the recovered GR parameters.

These findings motivate investigating 1PA modelling 
requirements in scalar-tensor theories of gravity, 
where an enlarged parameter space must be taken 
into account.

Such investigations are part of a broader research program that 
complements ongoing developments in GR, aiming to model 
EMRIs within effective field theory (EFT) frameworks 
featuring scalar fields non-minimally coupled to gravity. 
In such theories, the large mass asymmetry of these 
systems allows significant simplifications: at leading 
order in the mass ratio, 
deviations from the Kerr geometry of the primary can be 
neglected, while the scalar charge of the secondary 
controls departures from GR~\cite{Maselli:2020zgv}. Within the 
two-timescale approximation, the effects of the scalar 
charge can be incorporated modularly on top of a GR 
baseline model~\cite{Spiers:2023cva}.

At the adiabatic level, the leading scalar SF and 
associated fluxes have been computed for a variety of 
orbital configurations in the case of massless scalar fields~\cite{Barsanti:2022ana,Zhang:2022rfr,DellaRocca:2024pnm}, including fully 
generic orbits~\cite{Zi:2025lio,Gliorio:2026yvh}, as well as for equatorial circular orbits in the case of massive scalars~\cite{Barsanti:2022vvl}. These developments have enabled data analysis studies exploring the potential of EMRI observations to constrain the scalar charge and, more broadly, to probe the nature 
of the underlying scalar field~\cite{Maselli:2021men,Zhang:2022rfr,Barsanti:2022vvl,Speri:2024qak}.

While analyses performed so far have neglected subleading 
1PA contributions, recent work has begun to develop the 
framework required to compute 1PA scalar SF terms~\cite{Spiers:2023cva}. 
However, these contributions have not yet been incorporated 
into waveform models, owing to the technical complexity of the calculations.

Despite this limited progress, waveforms including only 
leading-order (adiabatic, or 0PA) scalar SF effects may 
still suffice for an unbiased recovery of the scalar charge.
In this work, we quantify the parameter-estimation bias introduced by neglecting 1PA gravitational and secondary spin effects in GW templates for EMRIs. To this end, we 
construct the first SF model that consistently combines 
a 1PA quasicircular Schwarzschild waveform in GR with 
leading-order scalar SF contributions.

We adopt an injection–recovery framework, generating 
synthetic LISA signals that include 1PA effects and analysing 
them with templates in which selected 0PA and 1PA contributions 
are systematically switched on or off. This setup allows us 
to characterize waveform systematics arising from model 
incompleteness beyond GR, by disentangling the impact of 
modelling inaccuracies in the GR and scalar sectors on 
parameter estimation, and in particular on the inference 
of the scalar charge.

Within this framework, we perform a Bayesian analysis based on Markov Chain 
Monte Carlo (MCMC) simulations to quantify the resulting parameter 
biases and the achievable constraints across the relevant 
parameter space.

\section{\label{sec:theoretical framework} Theoretical Framework}

We adopt the scalar-tensor SF model for non-spinning compact objects around massive black holes described in Refs.~\cite{Maselli:2020zgv, Spiers:2023cva}, 
which we briefly summarise here. We consider theories containing a real, massless, non-minimally coupled scalar field $\varphi$, governed by the action
\begin{equation}
S\left[\textbf{g}, \varphi, \Psi \right] = S_0\left[\textbf{g}, \varphi\right] + \alpha S_{\rm c} \left[\textbf{g}, \varphi\right] + S_{\rm m}\left[\textbf{g}, \varphi, \Psi\right]\,,
\label{action}
\end{equation}
where
\begin{equation}
S_0 = \int \dd ^4 x \frac{\sqrt{-g}}{16 \pi} \left(R - \frac{1}{2} \partial_a \varphi \partial^{a} \varphi \right)\,,
\end{equation}
$R$ is the Ricci scalar constructed from the metric tensor $\textbf{g}$, $S_{\rm c}$ is the interaction term describing the non-minimal coupling between the metric and the scalar field, and $\alpha$ is the fundamental coupling constant of the theory. We assume that $\alpha$ has dimensions of mass$^{n}$, with $n \geq 2$. Finally, $S_{\rm m}$ denotes the action of the matter fields $\Psi$. We also assume $S_{\rm c}$ to be analytic in $\varphi$.

We model EMRIs using relativistic perturbation theory, with the mass ratio $\epsilon = \mu/M \ll 1$ as the expansion parameter, where $M$ and $\mu$ denote the masses of the primary and secondary, respectively. Both the metric tensor and the scalar field are expanded order by order in $\epsilon$ as
\begin{align}
g_{ab} &= g^{(0)}_{ab} + \epsilon\,h^{(1)}_{ab} + \epsilon^2\, h^{(2)}_{ab} + \mathcal{O}(\epsilon^3)\,,\ \\
\varphi &= \varphi^{(0)} + \epsilon\, \varphi^{(1)} + \epsilon^2\,\varphi^{(2)} + \mathcal{O}(\epsilon^3)\,,
\end{align}
where $g^{(0)}_{ab}$ is the background metric of the primary, which we take here to be that of a non-rotating BH. 

In the absence of the secondary and for $\alpha = 0$, the exterior spacetime is a solution of the Einstein–Klein–Gordon equations and, by no-hair theorems~\cite{Sotiriou:2011dz,Hawking:1972qk}, is described by the Schwarzschild metric with a constant scalar field $\varphi^{(0)}$; hence, using shift symmetry, we can set $\varphi^{(0)}=0$. For $\alpha \neq 0$, assuming the  of BH solutions continuously connected to GR in the limit $\alpha \to 0$, deviations from Schwarzschild can be parametrized in terms of the dimensionless quantity
\begin{equation}
\zeta = \frac{\alpha}{M^n} = \epsilon^n \frac{\alpha}{\mu^n}\,.
\end{equation}
Current astrophysical constraints bound $\alpha/\mu^n$ to be $\sim \mathcal{O}(1)$, which implies $\zeta = \mathcal{O}(\epsilon^n) \ll 1$. As a result, corrections to the background metric and scalar field can be neglected at leading-order in the mass ratio, so that $\epsilon$ remains the sole perturbative bookkeeping parameter in our expansion.

The first-order perturbations $h_{ab}^{(1)}$ and $\varphi^{(1)}$ are sourced by the EMRI secondary, which is modelled using a skeletonised point-particle action with a scalar-dependent mass function $m[\tilde\varphi]$~\cite{Eardley:1975fgi,Damour:1992we,Julie:2017ucp,Julie:2017rpw}:
\begin{equation}
S_p = -\int_\gamma m[\tilde\varphi] \ d\tilde s\, ,
\end{equation}
where $\tilde g_{ab} = g_{ab} + h^{R}_{ab}$ and $\tilde\varphi = \varphi^{R}$ are the regular (effective) metric and scalar field, defined via a Detweiler–Whiting singular–regular decomposition~\cite{Detweiler:2002mi}. We expand the mass function in orders of the scalar field, giving
\begin{equation}
m[\tilde \varphi] = m^{(0)} + m^{(1)} \tilde\varphi + m^{(2)} \tilde\varphi^2 + \mathcal{O}(\tilde\varphi^3)\,.
\end{equation}
Using a buffer-region expansion of the first-order scalar field, 
\begin{equation}
\varphi^{(1)}
=
\frac{\mu d^{(0)}}{\hat r}
+\mathcal{O}\!\left(\frac{\mu^2}{\hat r^2}\right),
\qquad
\mu\ll\hat r\ll M,
\end{equation}
which defines the dimensionless scalar charge of the secondary $d^{(0)}$, one finds 
\begin{equation}
m^{(0)} = \mu, \qquad
m^{(1)} = -\frac{\mu d^{(0)}}{4}\,.
\end{equation}

At first order, the resulting field equations for the metric and scalar perturbations reduce to
\begin{align}
\delta G^{a b}[h^{(1)}_{a b}] &= 8 \pi \mu
\int \frac{\delta^{(4)}\left(x - y_p(\tau)\right)}{\sqrt{-g}}
\frac{\dd y^{a}_p}{\dd\tau}
\frac{\dd y^{b}_p}{\dd\tau} \, \dd\tau\,, \label{eq:einstein}\\
\Box \varphi^{(1)} &= - 4 \pi d \mu
\int \frac{\delta^{(4)} \left(x - y_p(\tau)\right)}{\sqrt{-g}} \, \dd\tau \,,
\label{eq:scalar}
\end{align}
where $\delta G_{ab}[\cdot]$ denotes the linearised 
Einstein operator about the background $g_{ab}$, 
and $\gamma$ is the worldline of the secondary, 
parametrised by proper time $\tau$ with four-velocity 
$u^a$.

Due to the decoupling of scales discussed above, at 
first order in $\epsilon$ the metric perturbation 
satisfies the standard GR field equations, while the 
scalar field obeys a wave equation sourced by a point 
charge in a Schwarzschild background.

At second order in the mass ratio, the metric perturbation $h^{(2)}_{ab}$ is governed by the linearised Einstein operator sourced by (i) the quadratic gravitational 
self-coupling $\delta^2 G_{ab}[h^{(1)}, h^{(1)}]$, 
(ii) the scalar stress--energy constructed from the 
first-order scalar field, and (iii) second-order 
corrections to the point-particle stress--energy 
arising from the use of effective (regular) fields 
in the particle action.

In this work, we neglect all first post-adiabatic contributions which involve scalar coupling, so that 
the second-order sector of the EMRI dynamics 
is effectively governed by GR contributions alone:
\begin{align}
\delta G_{ab}\!\left[h^{(2)}_{cd}\right] &=
-\delta^{2}G_{ab}\!\left[h^{(1)}_{cd},h^{(1)}_{cd}\right]+\delta G^{\langle1\rangle}\left[h^{(1)}_{ab}\right]
\nonumber \\  &+4\pi \!\int_{\gamma}\! \frac{\delta^{(4)}(x-z)}{\sqrt{-g}}\,
\mu\Big(
4 h^{R(1)}_{ac} u^{c} u_{b} \nonumber\\
& \ \ \ \ \ \ \ \ \ \ \ - u_{a}u_{b}\,(g_{cd}-u_{c}u_{d})\,h^{R(1)}_{cd}
\Big)\, d\tau \,.
\label{eq:second_order_metric_GR}
\end{align}
Here $\delta^{2}G_{ab}[\,\cdot\,,\,\cdot\,]$ 
denotes the quadratic contribution in the 
expansion of $G_{ab}$, while $\delta G^{\langle1\rangle}[\,\cdot\,]$ is the slow-time derivative 
linearised Einstein operator arising in a 
multiscale expansion~\cite{Miller:2020bft,Pound:2021qin}.

These simplifications allow us to leverage existing 
second-order results in GR~\cite{Pound:2019lzj,Warburton:2021kwk,Wardell:2021fyy} to construct a 1PA model in 
scalar-tensor theories of gravity.

The equation of motion for the secondary follows from varying the effective action, Eq.~\eqref{action}, with respect to the worldline coordinates:
\begin{equation}\label{eq:acceleration}
m[\tilde\varphi]\tilde a^a = m'[\tilde\varphi]\left(\tilde g^{ab} + \tilde u^a \tilde u^b\right)\nabla_b \tilde\varphi + \mathcal{O}(\epsilon^3)\,.
\end{equation}
Equation~\eqref{eq:acceleration} shows that the motion of the secondary deviates from a geodesic of the effective metric due to the presence of a scalar self-force. 

Within the two-timescale approximation, the dynamics 
of the secondary are described in terms of slowly 
evolving orbital elements, namely the semi-latus rectum $p$, the eccentricity $e$ and the inclination parameter $x_I$,  together with the orbital 
phases $\Phi_i=\{\Phi_r,\Phi_\theta,\Phi_\phi\}$. 
The phases evolve in time according to the corresponding orbital frequencies $\Omega_i=\{\Omega_r,\Omega_\theta,\Omega_\phi\}$ through $\dot{\Phi}_i = \Omega_i$. 
The orbital elements $(p,e,x_I)$ evolve slowly in time under 
radiation reaction, leading to a slow evolution of the 
frequencies $\Omega_i(p(t),e(t),x_I(t))$~\cite{Schmidt:2002qk,Chapman-Bird:2025xtd}.

Restricting to quasi-circular, equatorial inspirals 
sets $\Omega_r = \Omega_\theta = 0$ and $e=x_I=0$, removing the 
need to track $\Phi_r$ and $\Phi_\theta$. In this 
case, the orbital dynamics is described by the 
azimuthal phase $\Phi_\phi$, while the inspiral is 
driven by the slow evolution of $\Omega_\phi(r(t))$, where $r$ denotes the radius of the circular orbit. 
For details on the two-timescale formalism and the 
inclusion of slow-time-derivative contributions, see Refs.~\cite{Miller:2020bft, Pound:2021qin, Spiers:2023cva}. 
We neglect any slow-time derivative contribution involving the scalar field, as these are 1PA contributions that are further suppressed by taking the small coupling limit. Moreover, we also neglect the evolution of the mass and spin of the primary, a 1PA GR effect, which have been shown to be subdominant in 1PA analysis~\cite{Albertini:2022rfe, Martel:2003jj, Hughes:2018qxz}.

We further include an additional 1PA effect in our analysis, namely the spin of the secondary object, despite spin effects not being included in Ref.~\cite{Spiers:2023cva}. A complete treatment of spin within the scalar–tensor self-force framework is beyond the scope of this work. Instead, we adopt an approximate description in which spin effects are incorporated following their structure in GR~\cite{Harte:2011ku, Poisson:2011nh}. This approximation 
is justified under our assumption of small 
scalar coupling, whereby scalar contributions 
to spin effects are expected to enter at 1PA 
order or higher and are therefore consistently 
neglected here. A complete treatment of 
secondary spin within the scalar-tensor self-force framework will be presented in future work~\cite{jacopo}.  While the secondary spin is a small 1PA contribution, we include it in our analysis because the contribution has already been calculated and studied in 1PA binary models~\cite{Mathews:2021rod, Burke:2023lno}, offering an excellent test-bed for studying the necessity of including subdominant 1PA effects. 

Under this assumption, the 
spinning secondary can be treated as a test body 
following the Mathisson--Papapetrou--Dixon (MPD) equations~\cite{Mathisson:2010opl, Papapetrou:1951pa, Dixon:1970zza}, 
neglecting quadrupole and higher-order moments~\cite{Rahman:2026qho}. We 
follow Refs.~\cite{Mathews:2021rod, Burke:2023lno} 
for the incorporation of the MPD equations within 
the two-timescale framework.

To simplify our study of quasi-circular inspirals, 
we fix the orbital inclination to $\theta=\pi/2$, 
such that the orbital angular momentum is aligned 
with the $z$-axis. We further restrict the secondary 
spin to be (anti-)aligned with the $z$-axis. Under 
these conditions, both the orbital plane and the 
secondary spin do not precess. We define the 
dimensionless spin parameter as $\chi = S_2/\mu^2$, where $S_2$ is the spin angular momentum of the secondary. Positive (negative) values of $\chi$ corresponds to spin aligned (anti-aligned) 
with the orbital angular momentum.

\section{\label{sec:implementation} Implementation}

\subsection{Orbital Evolution}\label{sec:orbital_evolution}

To study the EMRI orbital dynamics, it is convenient 
to describe the evolution of quasi-circular 
trajectories in terms of the time evolution 
of the radial and azimuthal coordinates, 
using dimensionless variables $\hat t = t/M$ 
and $\hat r = r/M$. The dimensionless orbital 
frequency for circular orbits in Schwarzschild 
is then given by $\hat\Omega_\phi = \hat r^{-3/2}$.

The time evolution of $\hat r$ and $\Phi_\phi$ is 
governed by

\begin{equation}
\frac{d\hat{r}}{d\hat{t}} = - \nu\left[ F_0 (\hat{r}) + \nu F_1(\hat{r})\right]\,, \qquad 
\frac{d\Phi_\phi}{d\hat{t}} = \hat{\Omega}_\phi(\hat{r})\,,
\label{eq:rtevo}
\end{equation}
where the forcing functions are expanded in 
post-adiabatic orders. Equation~\eqref{eq:rtevo} 
is written using the symmetric mass ratio 
$\nu = M\mu/(M+\mu)^2$ as the expansion parameter~\cite{vandeMeent:2023ols}.

The adiabatic forcing term $F_0$ depends on the 
dissipative part of the first-order SF, including 
both gravitational and scalar contributions. The 
1PA correction $F_1$ includes contributions from 
the conservative part of the first-order SF and 
the dissipative part of the second-order SF, 
both computed within the gravitational sector.

Defining $\Lambda = d^2$, the forcing terms can 
be decomposed as
\begin{align}
F_0(\hat{r}) &= F^{\tn{GSF}}_0(\hat{r}) + \Lambda F^{\Lambda}_0(\hat{r})\,, \\
F_1(\hat{r}) &= F^{\tn{GSF}}_1(\hat{r}) + \chi F^{\chi}_1(\hat{r})\,,
\end{align}
where $F^{\tn{GSF}}_0$ and $F^{\tn{GSF}}_1$ denote 
the gravitational SF contributions, while $F^{\Lambda}_0$ 
and $F^{\chi}_1$ encode the effects of the scalar 
charge and the secondary spin, respectively.
The structure of the forcing terms directly determines 
the expansion of the accumulated orbital 
phase. Expanding perturbatively in $\nu$, and to linear 
order in $\Lambda$ and $\chi$, the orbital phase 
$\phi = \Phi_\phi$ can be written schematically as
\begin{align}
    \phi &= \frac{1}{\nu}\left(\phi^\tn{GSF}_0+ \Lambda \phi^\Lambda_0 \right) +  \phi^\tn{GSF}_1 + \chi \phi^\chi_1+ \nonumber\\
    &+  \mathcal{O}(\nu, \nu \Lambda, \nu \chi, \Lambda^2, \chi^2, \Lambda\chi)\,, 
\end{align}
where the leading-order phase contribution scales 
as $\nu^{-1}$ and contains both gravitational and 
scalar adiabatic effects, while the gravitational 
self-force and secondary spin corrections enter 
at first post-adiabatic order.

The four forcing terms can be related to the 
corresponding energy fluxes emitted by the binary 
as

\begin{equation}
\begin{aligned}
F_{0}^{A}
&= a(\hat r)\,\mathcal{F}_{0}^{A}\,, 
\qquad
\bigl(A \in \{\text{GSF},\,\Lambda\}\bigr)\,,\\
F_{1}^{A}
&= a(\hat r)\,\mathcal{F}_{1}^{A}
 + a(\hat r)^{2}\,
   \frac{\partial \hat E_{1}^{A}}{\partial \hat r}\,
   \mathcal{F}^{\text{GSF}}_{0}\,,
\qquad
\bigl(A \in \{\text{GSF},\,\chi\}\bigr)\,,
\label{eq:fluxes}
\end{aligned}
\end{equation}
where $a(\hat{r}) = (\partial \hat{E}_0 / \partial \hat{r})^{-1}$, $\hat{E}_0$ is the geodesic binding energy, 
$\hat{E}^{\chi}_1$ is the shift induced by the 
secondary spin~\cite{Jefremov:2015gza}, and 
$\hat{E}^\tn{GSF}_1$ is the correction induced 
by the conservative first-order SF~\cite{LeTiec:2011dp}.

While we used self-force results for $\hat{r} \in \left[6.25, 30\right]$,  for $\hat{r} \in \left(30, 60\right]$ we considered a hybrid trajectory model in which the post-adiabatic contributions are provided by PN expressions, as described in Appendix~\ref{appendix:hybrid model}.

In Fig.~\ref{fig:fluxes_comparison} we show the behaviour 
of the forcing terms $F^\Lambda_0$, $F^\tn{GSF}_1$, 
and $F^{\chi}_1$, normalized to the leading 
contribution $F^\tn{GSF}_0$, as functions of the 
orbital radius. We fix $\nu = 10^{-4}$, $\chi = 0.5$, 
and $\Lambda = 0.0025$ ($d = 0.05$). The figure shows 
that the scalar contribution grows at large radii 
and can become comparable to, or even dominate over, 
the first post-adiabatic  gravitational term for $r/M \gtrsim 8$, 
depending on the value of the scalar charge. 
The secondary spin contribution follows a behaviour 
similar to the first post-adiabatic gravitational term 
and remains subdominant over the range of orbital 
radii relevant for our analysis, becoming comparable 
only at smaller radii.
\begin{figure}[hbpt!]
\centering
\includegraphics[scale=0.55]{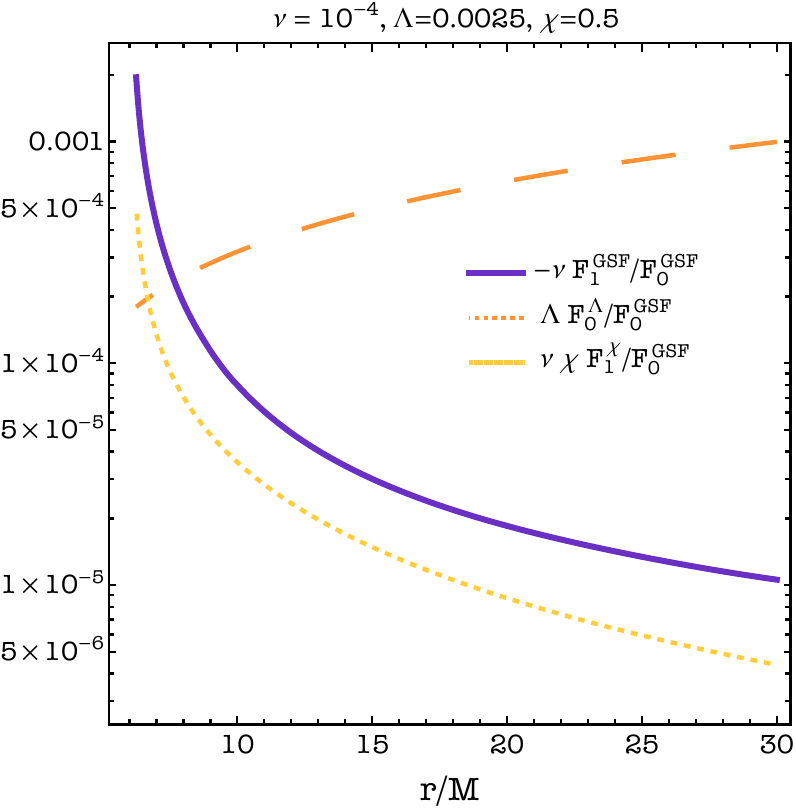}
\caption{Ratio between the sub-leading forcing 
terms $F^\tn{GSF}_1$, $F^\Lambda_0$, $F^\chi_1$ 
and the leading gravitational force $F^\tn{GSF}_0$ 
as a function of the orbital radius. The symmetric 
mass ratio is set to $\nu = 10^{-4}$, the scalar 
charge squared to $\Lambda = 0.0025$ and the secondary 
spin to $ \chi = 0.5$. }\label{fig:fluxes_comparison}
\end{figure}
%
\subsection{Waveform Models}\label{sec:waveformmodels}
We make use of the waveform model implemented in the 
recent release of FastEMRIWaveform (FEW)~\cite{Chapman-Bird:2025xtd, chapman_bird_2025_15630565}. For circular orbits in a 
Schwarzschild background, the gravitational 
waveform can be written as
\begin{equation}
    h(t) = h_+(t) - i h_\times(t) = \frac{\mu^\tn{red}}{D_S} \sum_{\ell m} H_{\ell m}(t,\theta,\phi)\, e^{-i \Phi_{m}(t)}\,,
\end{equation}
where $\mu^\tn{red} = \mu M /(\mu+M)$ is the 
reduced mass, $D_S$ is the luminosity distance 
to the source, and $(\theta,\phi)$ are the polar 
and azimuthal viewing angles in the source frame.

The mode amplitudes $H_{\ell m}$ are determined 
by the inhomogeneous solutions of the radial 
Teukolsky equation evaluated at $r \to +\infty$, 
together with the spherical harmonics. The phase 
is given by $\Phi_{m}(t) = m \Phi_\phi(t)$, with 
$d\Phi_\phi/dt = \hat{\Omega}_\phi$.

Using the \texttt{GenerateEMRIWaveform} class in 
FEW, we express the waveform, generated with the \texttt{FastKerrEccentricEquatorialFlux} model, in the solar system 
barycentric frame, introducing the angles 
$(\theta_S,\phi_S)$ and $(\theta_K,\phi_K)$, 
which specify the sky location of the source 
and the orientation of its orbital angular 
momentum, respectively.

The FEW infrastructure separates the waveform 
construction into two modules: a trajectory 
module and an amplitude module. We construct 
different waveform models depending on the 
underlying trajectory, and label them as 
1PAGR, 0PAGR, 1PAScalar, 0PAScalar,  1PASpinGR, 0PASpinGR, 1PASpinScalar, and 0PASpinScalar,
according to which forcing terms are 
included, as summarised in Table~\ref{tab:waveform_models}.

We neglect 1PA corrections to the waveform 
amplitudes, keeping only the leading-order (0PA) 
contribution in the amplitude module. As 
shown in~\cite{Burke:2023lno}, the impact 
of 1PA amplitude corrections on the waveform 
phase is negligible for signal-to-noise 
ratios $\ll 1/\epsilon$, a condition 
satisfied for the parameter space explored 
in this work.
\renewcommand{\arraystretch}{1.5}
\begin{table}[htbp!]
\centering
\begin{tabular}{cc}
\hline
 Waveform model & Forcing terms \\
\hline 
\hline
0PAGR& $F^\tn{GSF}_{0}$\\ 
1PAGR& $F^\tn{GSF}_{0}$, $F^\tn{GSF}_{1}$\\ 
0PAScalar& $F^\tn{GSF}_{0}$, $F^{\Lambda}_{0}$\\ 
1PAScalar&  $F^\tn{GSF}_{0}$, $F^{\Lambda}_{0}$,  $F^\tn{GSF}_{1}$\\ 
0PASpinGR& $F^\tn{GSF}_{0}$,   $F^\chi_{1}$\\ 
1PASpinGR& $F^\tn{GSF}_{0}$,   $F^\tn{GSF}_{1}$,$F^\chi_{1}$\\ 
0PASpinScalar& $F^\tn{GSF}_{0}$, $F^{\Lambda}_{0}$,  $F^\chi_{1}$\\ 
1PASpinScalar& $F^\tn{GSF}_{0}$, $F^{\Lambda}_{0}$,  $F^\tn{GSF}_{1}$,$F^\chi_{1}$\\ 
\hline
\hline
\end{tabular}
\caption{Definition of the waveform models used 
in this work. Each model is characterised by the 
set of forcing terms included in the orbital 
evolution, specifying whether gravitational (GSF), 
scalar ($\Lambda$), and spin ($\chi$) contributions 
are incorporated at adiabatic (0PA) and first 
post-adiabatic (1PA) order.}\label{tab:waveform_models}
\end{table}
\renewcommand{\arraystretch}{1.0}
%
\subsection{Data Analysis}
We perform the data analysis on simulated LISA data, 
obtained by passing the waveform model through 
the detector response. We employ the LISA response 
function implemented in~\cite{Katz:2022yqe}, 
assuming a static instrumental configuration 
with constant and equal arm lengths to generate 
the time-delay interferometry (TDI) $A$ and $E$ 
channels. The $T$ channel is neglected, as it is 
largely insensitive to GWs under the constant and equal-arm assumptions.

We carry out a fully Bayesian analysis through the use of Markov Chain Monte Carlo (MCMC). The 
posterior probability distribution $p(\bm{\theta}|d)$ 
of the waveform parameters $\bm{\theta}$, given the 
data $d$ (not to be confused with the secondary 
scalar charge), is obtained via Bayes' theorem, which 
relates the posterior to the likelihood $p(d|\bm{\theta})$ 
and the prior $p(\bm{\theta})$, up to a normalization 
constant:
\begin{equation}
\log p(\bm{\theta}|d) \propto \log p(d | \bm{\theta}) + \log p(\bm{\theta})\,.
\end{equation}

In our analysis, we use the \texttt{Eryn} sampler~\cite{michael_katz_2023_7705496}, based 
on \texttt{emcee}~\cite{emcee}, which provides 
additional features such as parallel tempering. 
In this work, we adopt a single temperature 
setup and use 50 walkers for all MCMC simulations. 
Since our goal is not to perform a blind search 
for EMRI signals, we initialise the chains in 
the vicinity of the true parameters and evolve 
them until convergence is achieved.

The data stream $d^{(X)}$ in a given TDI channel 
$X$ is modelled as
\begin{equation}
d^{(X)}(t) = h^{(X)}_e(t;\bm{\theta}_{\tn{true}}) + n^{(X)}(t)\,, \qquad X=\{A,E\}\,,
\end{equation}
where $h^{(X)}_e$ denotes the exact waveform and 
$n^{(X)}$ the instrumental noise, assumed to be 
Gaussian and stationary. Under the assumption of 
equal and constant arm lengths, the $A$ and $E$ 
channels are uncorrelated, facilitating the use of a purely diagonal noise covariance over $A$ and $E$.

In this work, we neglect instrumental noise, 
following the arguments presented in~\cite{Burke:2023lno}, 
and adopt the Gaussian likelihood~\cite{whittle:1957}
\begin{equation}
 \log p(d|\bm{\theta}) \propto - \frac{1}{2} \sum_{X=\{A,E\}} \langle h_e - h_m \,|\, h_e - h_m \rangle_{(X)}\,,
\end{equation}
where $\langle \cdot \,|\, \cdot \rangle$ denotes 
the noise-weighted inner product between two 
waveforms, defined as~\cite{Finn:1992wt}
\begin{equation}
\langle h_1 \vert h_2 \rangle = 4\,\Re \int_{0}^{\infty} \frac{\tilde{h}_1(f)\,\tilde{h}_2^{\star}(f)}{S_n(f)}\, \dd f\,,
\label{eq:sca_prod}
\end{equation}
where $\tilde{h}(f)$ denotes the Fourier transform 
of the time-domain waveform $h(t)$. The noise power 
spectral density $S_n(f)$ is taken from the LISA 
sensitivity curve; in particular, we adopt the 
SciRDv1 model~\cite{LISAsr:18aa} for the $(A,E)$ 
channels, including the white-dwarf confusion noise.
The signal-to-noise ratio (SNR) of a signal $h_1$ 
is defined as $\rho = \mathrm{SNR} = \langle h_1 \vert h_1 \rangle^{1/2}\,.$ 

The parameter vector is given by $\bm{\theta} = \{M, \mu, \hat{r}_0, D_S, \theta_S, \phi_S, \theta_K, \phi_K, \Phi_{\phi_0}, \Lambda,\chi \}$. 
The masses $M$ and $\mu$ are detector-frame quantities, 
related to the source-frame masses by 
$M = (1 + z) M_{\text{s}}$ and $\mu = (1 + z) \mu_{\text{s}}$.

For all analyses, we inject the angular parameters 
$(\theta_S, \phi_S, \theta_K, \phi_K, \Phi_{\phi_0}) = (0.8, 2.2, 1.6, 1.2, 2.5)$ 
radians and consider an observation time of $T_\tn{obs} = 1$ year. 
The luminosity distance $D_S$ is determined by fixing 
the SNR of the injected waveform. We include all 
available amplitude modes in both the injection and 
recovery, setting the FEW parameter 
\texttt{mode\_selection\_threshold} = 0.0.

Finally, we adopt uniform priors for all parameters, 
centred on the injected values $\bm{\theta}_\tn{tr}$ 
and chosen to be sufficiently broad so as not to 
constrain the posterior. 
In practice, the prior widths are selected empirically from preliminary exploratory runs to ensure that the sampled chains remain well contained within the prior boundaries throughout the analysis.
Explicitly, for each parameter $\theta^i$ we assume
\begin{equation}
\theta^i \sim \theta^{i}_\tn{tr} + 25\, U\big[-\Delta\theta^i, \Delta\theta^i\big]\,,
\end{equation}
where $U[a,b]$ denotes a uniform distribution over the 
interval $[a,b]$, and 
$\Delta\theta^i$ sets the characteristic scale of the parameter exploration around the injected value. For the secondary spin parameter, we instead adopt a uniform prior over the interval $\chi \in [-1,1]$.

\section{\label{sec:results} Results}
%
\renewcommand{\arraystretch}{1.5}
\begin{table*}[htbp]
\centering
\begin{tabular}{ccccccccccccc}
\hline
 $\epsilon$ & Fig. & SNR & $d^{(\textnormal{inj})}$& Injected Waveform &Recovered Waveform & $\Delta\Phi_\phi $& $\log \mathcal{L}^{\textnormal{(inj)}} $ & $\log \mathcal{L}^{\textnormal{(bf)}} $ & $\frac{\Delta\theta_{M}}{\sigma_M}$ & $\frac{\Delta\theta_{\mu}}{\sigma_\mu}$ & $\frac{\Delta\theta_{\hat{r}_0}}{\sigma_{\hat{r}_0}}$ & 
 $\frac{\Delta\theta_{\Lambda}}{\sigma_{\Lambda}}$ \\ 
\hline 
\hline 
$5\times10^{-5}$ & \ref{fig:2}& 200 & 0.05 &1PAScalar &0PAScalar  & $-7$ &$- 40087$  & $-0.498$ & $5.2$ & $0.6$ & $5.7$ & $0.07$\\ 
$10^{-4}$&  \ref{fig:2}&
200 &0.05&1PAScalar&0PAScalar & $-65$ &$- 36292$ & $-0.187$ &$14.5$ & $5.4$ & $14.9$ & $0.3$\\
$10^{-4}$&  \ref{fig:2}&
200 &1.00&1PAScalar&0PAScalar & $-50$ &$- 37920$ & $-0.304$ &$13.0$ & $4.8$ & $13.4$ & $1.7$\\
$4\times10^{-4}$&  \ref{fig:2}&
100&0.05&1PAScalar&0PAScalar & $-148$ &$-9622$ & $-0.243$ & $3.8$ & $0.2$ & $3.9$ & $2.2$\\
\hline
$10^{-4}$&  \ref{fig:3}&50 &0.025&1PAScalar&1PAGR & 507 &$-2375$ & $-14.528$& $15.5$ & $30.5$ & $14.5$ & - \\
$10^{-4}$&\ref{fig:3} &50 &0.025&1PAScalar&0PAGR & 446 &$-2383$ & $-14.047$ & $26.3$ &$37.5$ &$24.9$ & -\\
\hline
$10^{-4}$& \ref{fig:4}  &50 &0.05&1PASpinScalar&0PASpinScalar & 66 &$-2264$& $ -0.241$ &$2.4$ &$1.2$&$2.4$& $0.2$ \\
\hline
$10^{-4}$& \ref{fig:6} & 50 & 0.05&0PAScalar & $-$1PNScalar & 107& $-2333$ & $-0.219$ & $0.3$ & $0.5$ & $0.3$ & $0.5$\\
\hline
\hline
\end{tabular}
\caption{Summary of the simulation results. The table reports 
the orbital dephasing between the injected and recovered waveforms, 
both evolved from the same initial conditions, the log-likelihood 
evaluated at the injected and best-fit parameters, and the parameter 
biases—expressed in units of the corresponding statistical 
uncertainties—for the intrinsic binary parameters $M$, $\mu$, 
$\hat{r}_0$, and $\Lambda$. 
}
\label{tab:log_likelihood}
\end{table*}
\renewcommand{\arraystretch}{1.0}
To assess the impact of post-adiabatic corrections on the 
measurement of the scalar charge, we consider four 
injection–recovery scenarios:
\begin{itemize}
\item Case I: injection with the 1PAScalar model and recovery with the 0PAScalar model (Section~\ref{sec:results:1PAScalarVS0PAScalar}). This setup tests biases in parameter inference arising from neglecting 1PA gravitational self-force corrections. 
\item Case II: injection with the 1PAScalar model and recovery with the 1PAGR and 0PAGR models (Section~\ref{sec:results:1PAScalarVSGR}). Here, we investigate whether scalar-induced deviations can be misinterpreted within GR, and evaluate the resulting biases when scalar effects are neglected.
\item Case III: injection with the 1PASpinScalar model and recovery with the 0PASpinScalar model (Section~\ref{sec:results:1PASpinScalarVS0PASpinScalar}). In this case we include the secondary spin to assess its impact and quantify biases arising from neglecting 1PA gravitational self-force corrections while retaining 1PA spin contributions. 
\item Case IV: injection with the 0PAScalar model and recovery with the $-1$PNScalar model, defined in Section~\ref{sec:results:0PAScalarVS-1PNScalar}. This case assesses the validity of modelling scalar emission using a leading-order dipole ($-1$PN) approximation, and quantifies the associated systematic errors relative to a fully relativistic treatment.
\end{itemize}
A summary of the results for these four cases is presented in Table~\ref{tab:log_likelihood}. We report the orbital dephasing 
\begin{equation}
\Delta\Phi_\phi \equiv \Phi^{\tn{(inj)}}_{\phi,\tn{fin}} - \Phi^{\tn{(rec)}}_{\phi,\tn{fin}}
\end{equation} computed by evolving the injection and 
recovery waveform models from the same injected initial conditions up to the end of the observation time $T_\tn{obs}$. We also report the log-likelihood evaluated at the injected 
parameters, $\log\mathcal{L}^{\tn{(inj)}}$, and at the 
best-fit parameters, $\log\mathcal{L}^{\tn{(bf)}}$, as well 
as the parameter biases, expressed in units of the 
corresponding statistical uncertainties, for the intrinsic 
binary parameters $M$, $\mu$, $\hat{r}_0$, and $\Lambda$. 
As one can see, convergence in all simulations has been reached 
with high best-fit log-likelihood compared to the injection log-likelihood. 
\subsection{\label{sec:results:1PAScalarVS0PAScalar} Case I: 1PAScalar vs 0PAScalar}
\begin{figure*}[htbp!]
\centering
\includegraphics[width=0.49\textwidth]{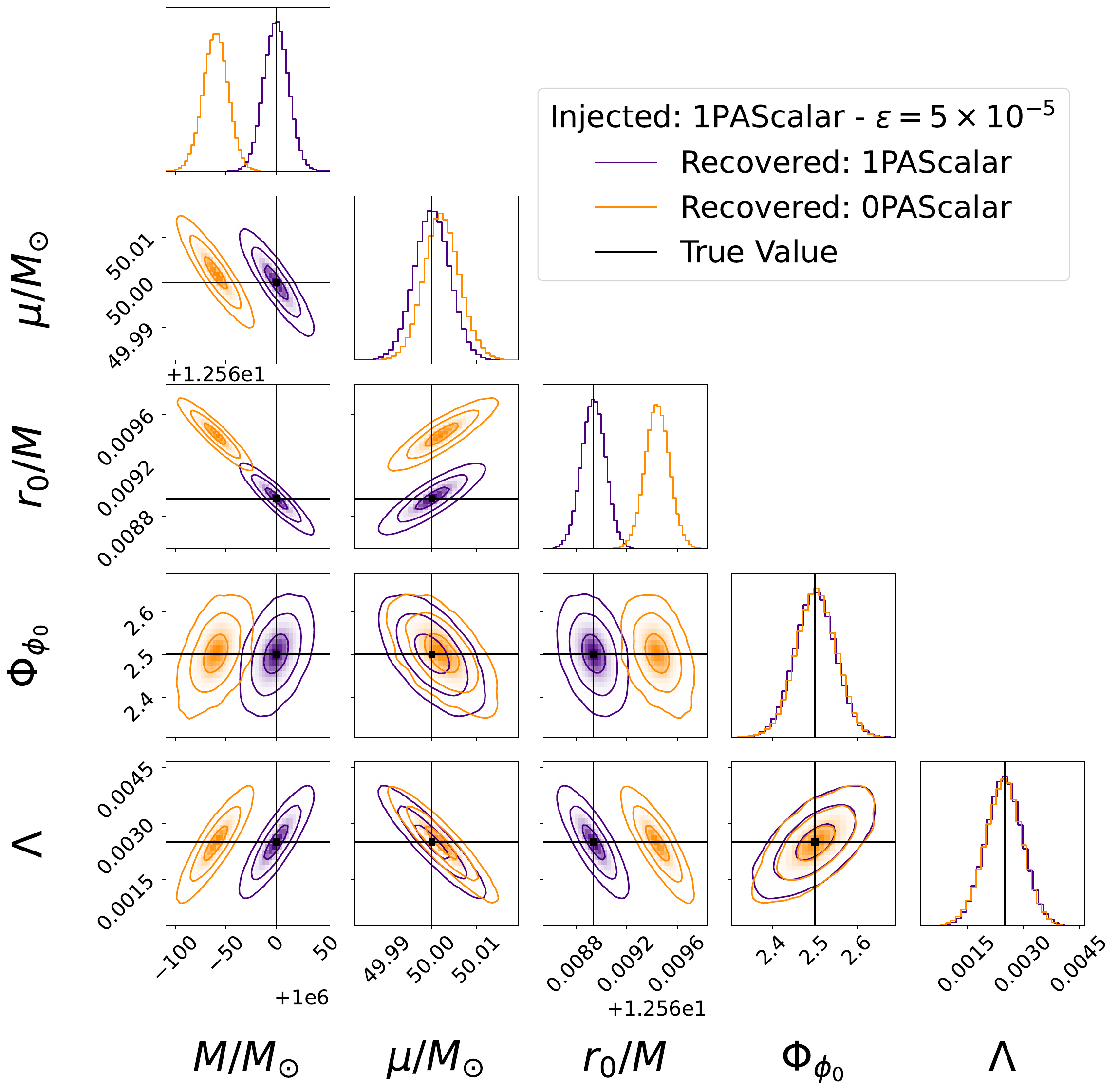}
\hfill
\includegraphics[width=0.49\textwidth]{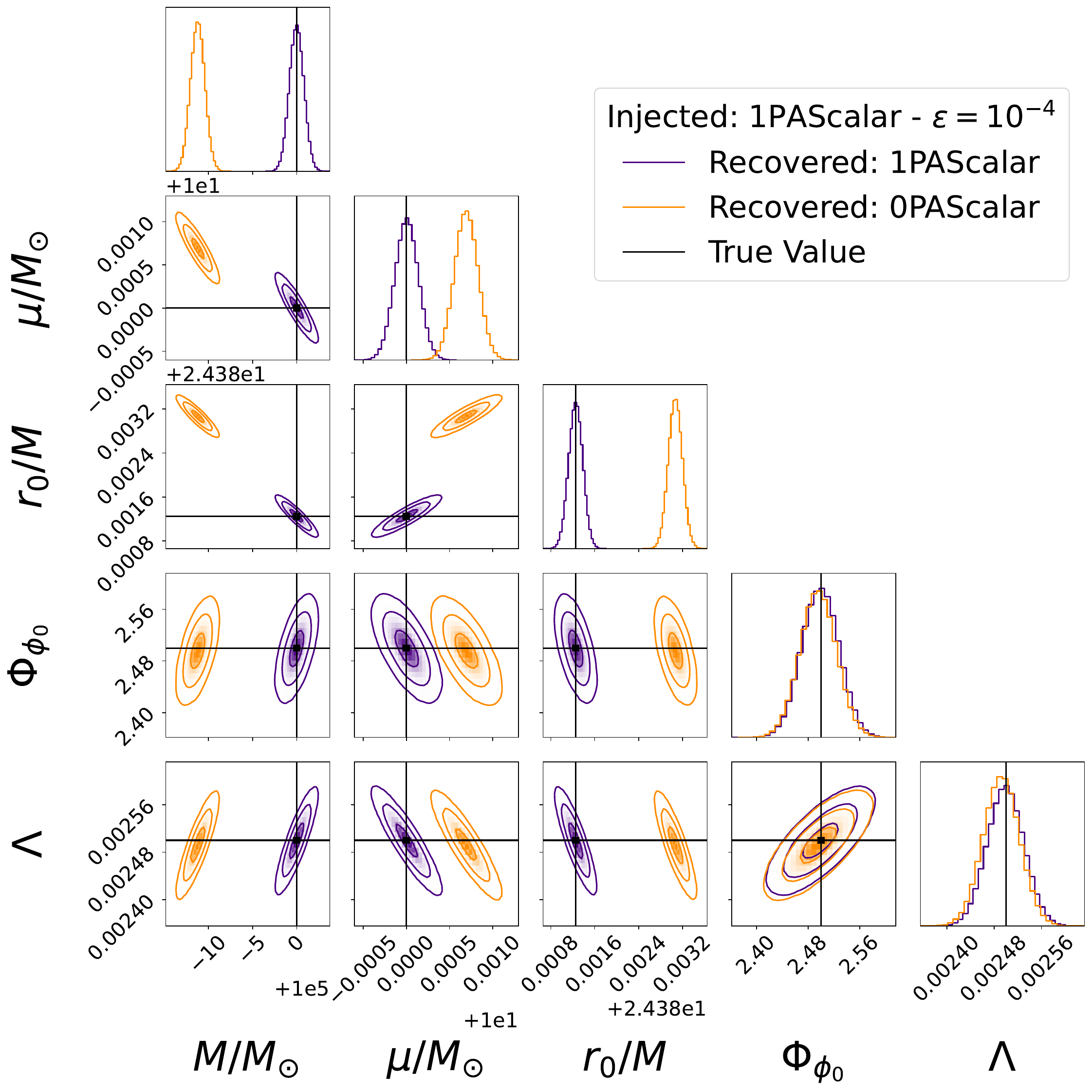}
\vspace{0.2cm}
\includegraphics[width=0.49\textwidth]{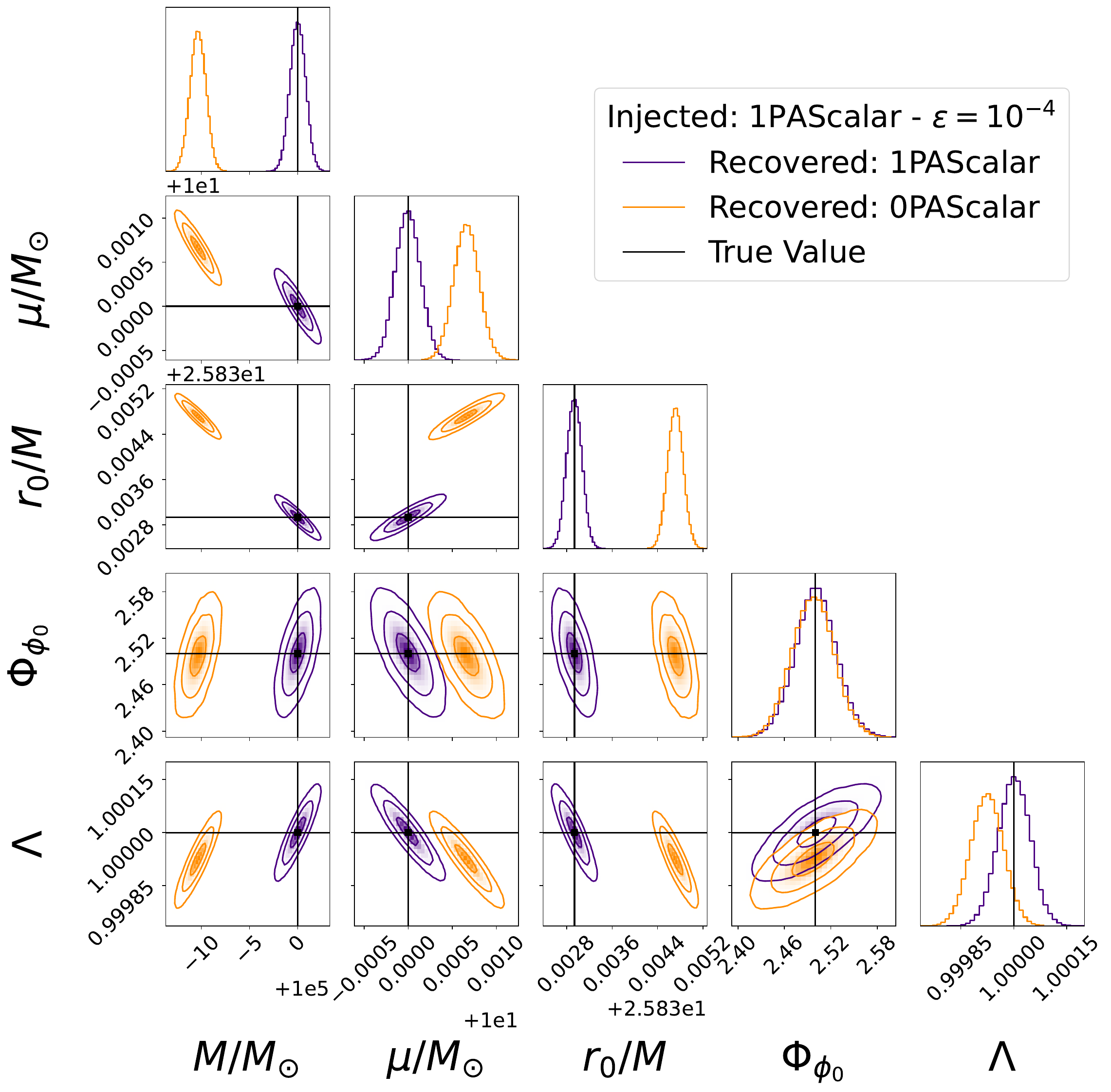}
\hfill
\includegraphics[width=0.49\textwidth]{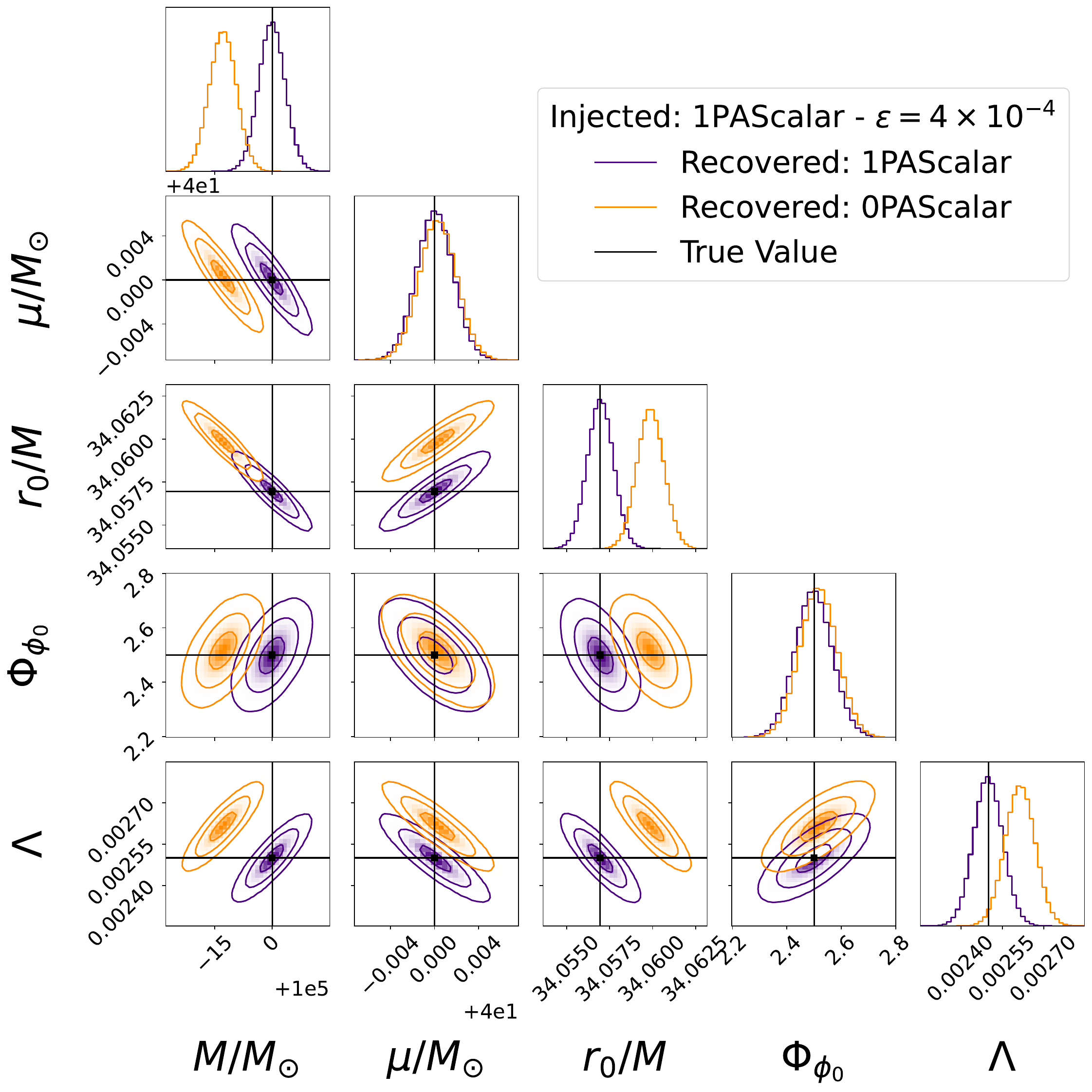}
\caption{
Corner plots of the intrinsic parameters, showing the posterior distributions obtained by injecting the 1PAScalar 
model and recovering with the 1PAScalar (purple) and 0PAScalar (orange) 
models. The panels correspond to: 
top-left, $\epsilon = 5\times10^{-5}$, $d=0.05$, and $\tn{SNR}=200$; 
top-right, $\epsilon = 10^{-4}$, $d=0.05$, and $\tn{SNR}=200$; 
bottom-left, $\epsilon = 10^{-4}$, $d=1$, and $\tn{SNR}=200$; 
bottom-right, $\epsilon = 4\times10^{-4}$, $d=0.05$, and $\tn{SNR}=100$.
}
\label{fig:2}
\end{figure*}
We first investigate waveform systematics between the 0PAScalar 
and 1PAScalar models. We consider three representative mass 
ratios, $\epsilon = 5\times10^{-5},\,10^{-4},\,4\times10^{-4}$ with injected scalar charge $d=0.05$. For $\epsilon = 10^{-4}$, we also consider a larger scalar charge, $d=1$.  
Figure~\ref{fig:2} shows the four corresponding corner plots 
for the posterior distributions of the intrinsic binary 
parameters, with purple (orange) contours corresponding 
to recovery with the correct (mismodelled) waveform. Corner 
plots including all ten parameters are provided in Appendix~\ref{appendix:corner plots}, Figs.~\ref{appendix:fig:7},~\ref{appendix:fig:8},~\ref{appendix:fig:9}, and~\ref{appendix:fig:10}.

The top-left panel of Fig.~\ref{fig:2} corresponds to 
a system with mass ratio $\epsilon = 5\times10^{-5}$, 
primary mass $M = 10^6\,M_\odot$, and secondary mass $\mu = 50\,M_\odot$. The inspiral starts at an initial orbital 
radius $\hat{r}_\tn{0} \simeq 12.57$, with scalar 
charge $d = 0.05$ ($\Lambda = 0.0025$). The final 
orbital radius is $\hat{r}^{\tn{0PAScalar}}_{\tn{fin}} \simeq 6.35$ and $\hat{r}^{\tn{1PAScalar}}_{\tn{fin}} \simeq 6.36$ 
for the 0PAScalar and 1PAScalar trajectories, respectively. 
The resulting orbital dephasing between the two trajectories 
is
\[
\Delta \Phi_\phi = \Phi^{\tn{1PAScalar}}_{\phi,\tn{fin}} - \Phi^{\tn{0PAScalar}}_{\phi,\tn{fin}} \simeq -7 \ \text{rad}.
\]
The SNR of the injected 1PAScalar waveform is fixed 
to 200.

When using the 0PAScalar template to recover the signal, 
the parameters most affected are the primary mass and the initial orbital radius, both showing biases of 
$\simeq 5\sigma$, and the secondary mass $\mu$, 
with a bias of $\simeq 0.6\sigma$. All other 
parameters remain essentially unbiased, with 
deviations $\lesssim 0.1\sigma$.

Remarkably, even at such a high SNR ($\tn{SNR} = 200$), 
no significant bias is observed in the scalar charge. 
This suggests that the intrinsic parameters 
$(M, \mu, \hat{r}_0)$ alone can compensate 
for the differences between the two models 
for this binary configuration.

A similar behaviour is observed for a larger mass ratio, 
$\epsilon = 10^{-4}$, shown in the top-right panel. This configuration 
is analogous to that of the top-left panel, 
with the same scalar charge but a primary mass 
$M = 10^5\,M_\odot$ and a secondary mass $\mu = 10\,M_\odot$. 
The inspiral starts at $\hat{r}_0 \simeq 24.38$, leading, 
after one year, to final radii $\hat{r}^{\tn{0PAScalar}}_{\tn{fin}} \simeq 6.36$ and $\hat{r}^{\tn{1PAScalar}}_{\tn{fin}} \simeq 6.41$ 
for the 0PAScalar and 1PAScalar trajectories, respectively. 
The corresponding dephasing is $\Delta \Phi_\phi \simeq -65$ rad. 
The SNR of the injected 1PAScalar waveform is fixed to 200.

When recovering with the 0PAScalar template, we find larger 
biases than in the previous case: $\simeq 15\sigma$ for $M$ 
and $\hat{r}_0$, and $\simeq 5\sigma$ for $\mu$. All 
other parameters show biases $\lesssim 0.3\sigma$, 
including the scalar charge, which again remains unbiased.

The bottom-left panel shows a configuration with the same component masses, 
($\epsilon = 10^{-4}$), but with injected scalar charge  $d=1$. The initial orbital radius is $\hat{r}_0 \simeq 25.83$, leading to $\hat{r}^\tn{1PAScalar}_\tn{fin} \simeq 6.39$, $\hat{r}^\tn{0PAScalar}_\tn{fin} \simeq 6.35$ and a corresponding dephasing $\Delta \Phi_\phi \simeq 50$ rad after 1 year of evolution. In this case, when recovering with the 0PAScalar model, we observe a bias of $\simeq 13\sigma$ in $M$ and $\hat{r}_0$, $\simeq 5\sigma$ in $\mu$, and $\simeq 1.7\sigma$ in $\Lambda$. All other parameters show biases $\lesssim 0.3\sigma$. Unlike the lower-charge configurations, the scalar charge is no 
longer recovered without significant bias. This indicates that, for 
larger scalar couplings, the mismodelling induced by neglecting 
the 1PA gravitational corrections cannot be entirely absorbed by 
the orbital parameters alone, and partially propagates into the 
scalar sector.

Finally, the bottom-right panel shows 
the corner plot for the largest mass ratio considered, 
$\epsilon = 4 \times 10^{-4}$. The binary has component 
masses $M = 10^5\,M_\odot$ and $\mu = 40\,M_\odot$. The 
inspiral starts at $\hat{r}_0 \simeq 34.06$ and reaches 
final radii $\hat{r}^{\tn{0PAScalar}}_{\tn{fin}} \simeq 6.35$ 
and $\hat{r}^{\tn{1PAScalar}}_{\tn{fin}} \simeq 6.65$, 
with a corresponding dephasing of $\Delta \Phi_\phi \simeq -148$ rad. In this case, the injected signal has $\tn{SNR} = 100$, 
as higher values (e.g.\ $\tn{SNR} = 200$) lead to 
convergence issues in the recovery.

The primary mass and initial radius are biased by 
$\simeq 1.5\sigma$. This is the only configuration with injected charge $d=0.05$
that shows a noticeable bias in the scalar charge, 
$\Delta \theta_\Lambda \simeq 2.2\,\sigma_\Lambda$. 

A bias in $\Lambda$ can be mapped onto a bias in the 
fundamental parameters of a given theory of gravity, 
once the relation between the scalar charge and the 
underlying coupling is specified. A well-studied example 
is scalar Gauss–Bonnet (sGB) gravity, in which the 
coupling constant $\alpha_\tn{sGB}$ can be expressed 
in terms of $\Lambda$ as~\cite{Julie:2019sab}
\begin{equation}
\sqrt{\alpha_\tn{sGB}} = \sqrt{2}\, \mu \,\Lambda^{1/4}\,.
\end{equation}
The bias on $\sqrt{\alpha_\tn{sGB}}$ is obtained by 
propagating the joint posterior distribution of the 
sampled parameters $(\Lambda, \mu)$. We find 
$\sqrt{\alpha^{\tn{1PA}}_\tn{sGB}} \simeq 12.65 \pm 0.065\,\text{km}$ 
and $\sqrt{\alpha^{\tn{0PA}}_\tn{sGB}} \simeq 12.79 \pm 0.063\,\text{km}$, 
corresponding to a bias of $\simeq 2\sigma$.

Overall, the results show that neglecting 1PA gravitational corrections biases the intrinsic orbital parameters, particularly $M$ and $\hat{r}_0$, while leaving the scalar charge largely unaffected over a broad range of mass ratios. A significant bias in $\Lambda$ appears only at the largest $\epsilon$ or for sufficiently large scalar charges.
\subsection{\label{sec:results:1PAScalarVSGR}Case II: 1PAScalar vs 1PAGR and 0PAGR}
\begin{figure*}[htbp!]
\centering
\includegraphics[scale=0.26]{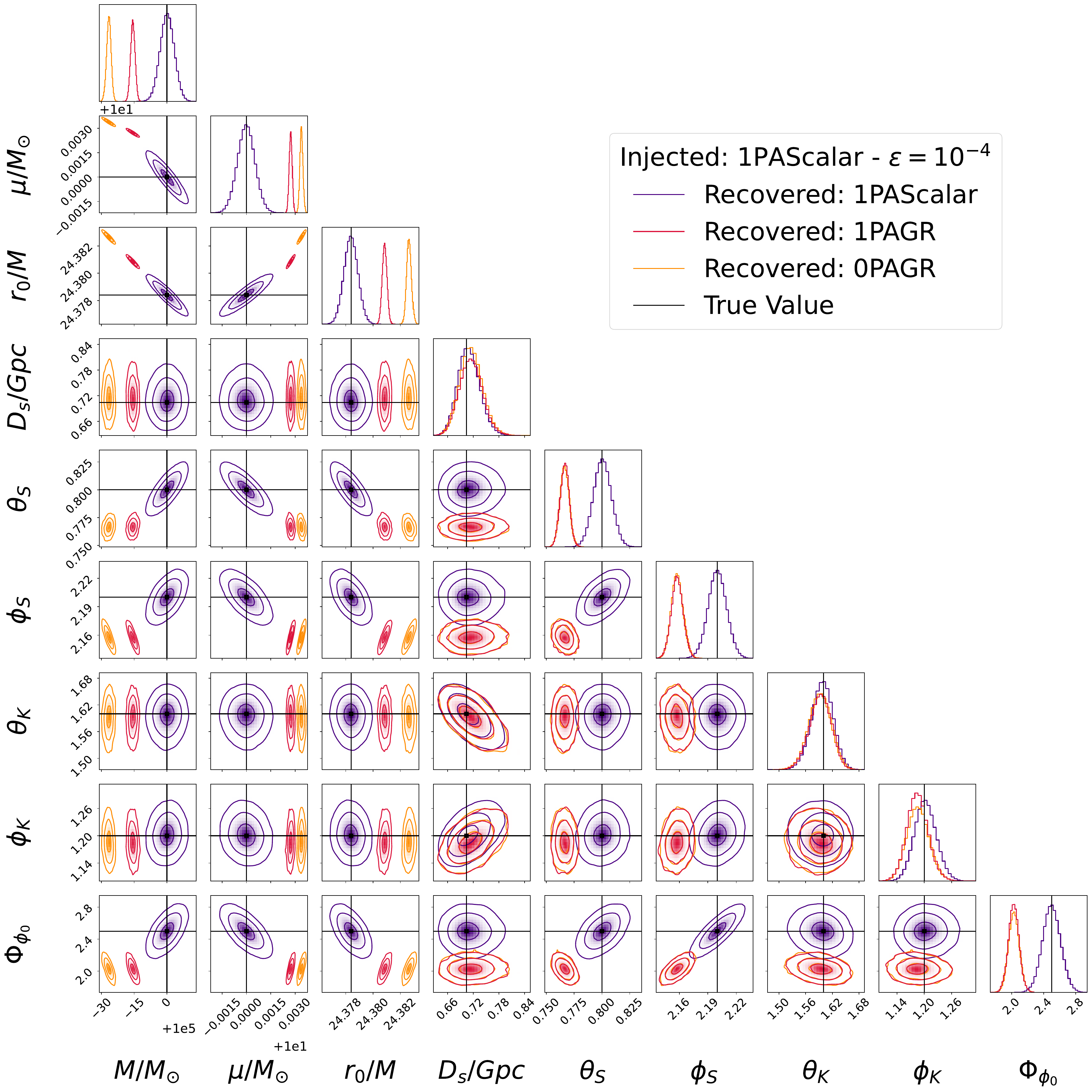}
\caption{Corner plot of the binary parameters, obtained 
by injecting a 1PAScalar waveform and recovering with 
the 1PAScalar (indigo), 1PAGR (red), and 0PAGR (orange) 
models. The injected scalar charge is $d = 0.025$ 
($\Lambda = 0.000625$). The injected signal has 
$\tn{SNR} = 50$.
}\label{fig:3}
\end{figure*}
We next investigate recovery with pure-GR templates, 
considering an injected 1PAScalar EMRI inspiral with initial 
radius $\hat{r}_0 \simeq 24.38$ and scalar charge 
$d = 0.025$ ($\Lambda = 0.00625$), such that the 
final radius is $\hat{r}^{\tn{1PAScalar}}_{\tn{fin}} \simeq 6.41$. 
We perform the analysis using both 1PAGR and 0PAGR 
templates for recovery. The results are summarised 
in Fig.~\ref{fig:3}, which shows the 
corner plot for all ten parameters, and in Table~\ref{tab:biases_GR_recovery}, which reports 
the corresponding parameter biases.
\begin{table}[htbp!]
\centering
\begin{tabular}{ccc}
\toprule
 $\theta$ & $\frac{\Delta \theta}{\sigma}$ (0PAGR)&  $\frac{\Delta \theta}{\sigma}$ (1PAGR)\\
\midrule
\hline
$M$                 & 26.2 & 15.5 \\
$\mu$             & 37.5 & 30.5 \\
$\hat{r}_0$         & 24.8 & 14.5 \\
$D_S $                 & 0.4  & 0.4  \\
$\theta_S$ & 8.4 & 8.6 \\
$\phi_S$     & 7.0 & 7.0 \\
$\theta_K$ & 0.3 & 0.3 \\
$\phi_K$     & 0.6 & 0.6 \\
$\Phi_{\phi_0}$ & 7.2 & 7.5 \\
\hline
\hline
\end{tabular}
\caption{Parameters biases in units of posterior standard deviation for the recovery with the 0PAGR (second column) and 1PAGR (third column) models.}\label{tab:biases_GR_recovery}
\end{table}
We find significant biases in the intrinsic parameters 
$M$, $\mu$, $\hat{r}_0$, and $\Phi_{\phi_0}$, as well 
as in the extrinsic angles $\theta_S$ and $\phi_S$. 
These parameters exhibit strong correlations with the 
scalar charge, which is absent in the recovery model. 
A comparison between the 1PAGR and 0PAGR recoveries 
shows that the largest differences arise in the 
intrinsic parameters $M$, $\mu$, and $\hat{r}_0$, 
in agreement with the findings of the previous section.

We also find that neglecting the scalar charge in the 
recovery leads to overly optimistic estimates of 
parameter uncertainties, as correlations with the 
additional degree of freedom are not properly 
captured. This highlights the risk of underestimating 
uncertainties when using incomplete waveform models, 
and reinforces the importance of including such effects in the modelling.

In summary, neglecting the scalar emission in the recovery leads to large biases in both intrinsic parameters and extrinsic parameters, as the missing scalar degree of freedom is absorbed through parameter correlations. At the same time, uncertainties are underestimated due to neglecting correlations between the vacuum--GR parameters and the secondary flux -- highlighting the limitations of incomplete waveform models. 
\subsection{\label{sec:results:1PASpinScalarVS0PASpinScalar}Case III: 1PASpinScalar vs 0PASpinScalar}
In Case III we extend the analysis to include 
the spin of the secondary as an additional 
parameter entering the waveform at post-adiabatic 
order. In this setup, signals are generated using 
the 1PASpinScalar model and analysed with the 
0PASpinScalar template.

The corner plot for the intrinsic parameters 
$(M, \mu, \hat{r}_0, \Lambda, \chi)$ is shown 
in Fig.~\ref{fig:4}, while the full 
posterior distributions are presented in Appendix~\ref{appendix:corner plots}, Fig.~\ref{appendix:fig:11}. The injected 
parameters are the same as in Fig.~\ref{fig:3}, with the addition of a secondary spin $\chi = 0.5$. 
This configuration yields final radii $\hat{r}^{\tn{1PASpinScalar}}_{\tn{fin}} \simeq 6.40$ 
and $\hat{r}^{\tn{0PASpinScalar}}_{\tn{fin}} \simeq 6.35$ 
for the 1PASpinScalar and 0PASpinScalar trajectories, 
respectively.
We find that the secondary spin is not constrained, with 
the marginal posterior on $\chi$ remaining uninformative 
with respect to its prior. 
Moreover, we observe biases of $\simeq 2.4\sigma$ in $M$ 
and $\hat{r}_0$, and $\simeq 1.2\sigma$ in $\mu$, while 
all other parameters exhibit biases $\lesssim 0.3\sigma$.

We can compare our result with those reported in~\cite{Burke:2023lno} (see Fig. 4), in which, for the same mass ratio $\epsilon = 10^{-4}$, the secondary spin was found to be constrained. However, when repeating the same analysis using the latest \texttt{FastKerrEccentricEquatorialFlux} waveform model, the secondary spin is no longer constrained. The details of such an analysis are provided in Appendix~\ref{appendix:secondary_spin}, with the corresponding posterior distributions  shown in the corner plot in Fig.~\ref{appendix:fig:13}. We believe that this discrepancy originates from a bug in the \texttt{FastSchwarzschildEccentricFlux} waveform model, which introduced a higher-mode content in the waveform, and that the results presented here should be therefore considered as the reliable ones. 
We do remark that the conclusions of ~\cite{Burke:2023lno} remain unchanged -- verified through this paper -- that the 1PA components of the gravitational SF are essential for parameter estimation, further validated through this work. For completeness, we also investigated the case with the largest mass ratio considered in~\cite{Burke:2023lno}, $\epsilon = 10^{-3}$, and performed an additional recovery study in which the secondary spin was neglected in the template waveform. The results are qualitatively similar to ~\cite{Burke:2023lno} with weak posterior support on the secondary spin parameter. Further Details are provided in Appendix~\ref{appendix:secondary_spin}.

Overall, the inclusion of the secondary spin has a negligible impact on the scalar charge inference, while the secondary spin parameter itself remains unconstrained in the analysis considered here, even in the pure-GR case. 
\begin{figure}[hbpt!]
\centering
\includegraphics[scale=0.225]{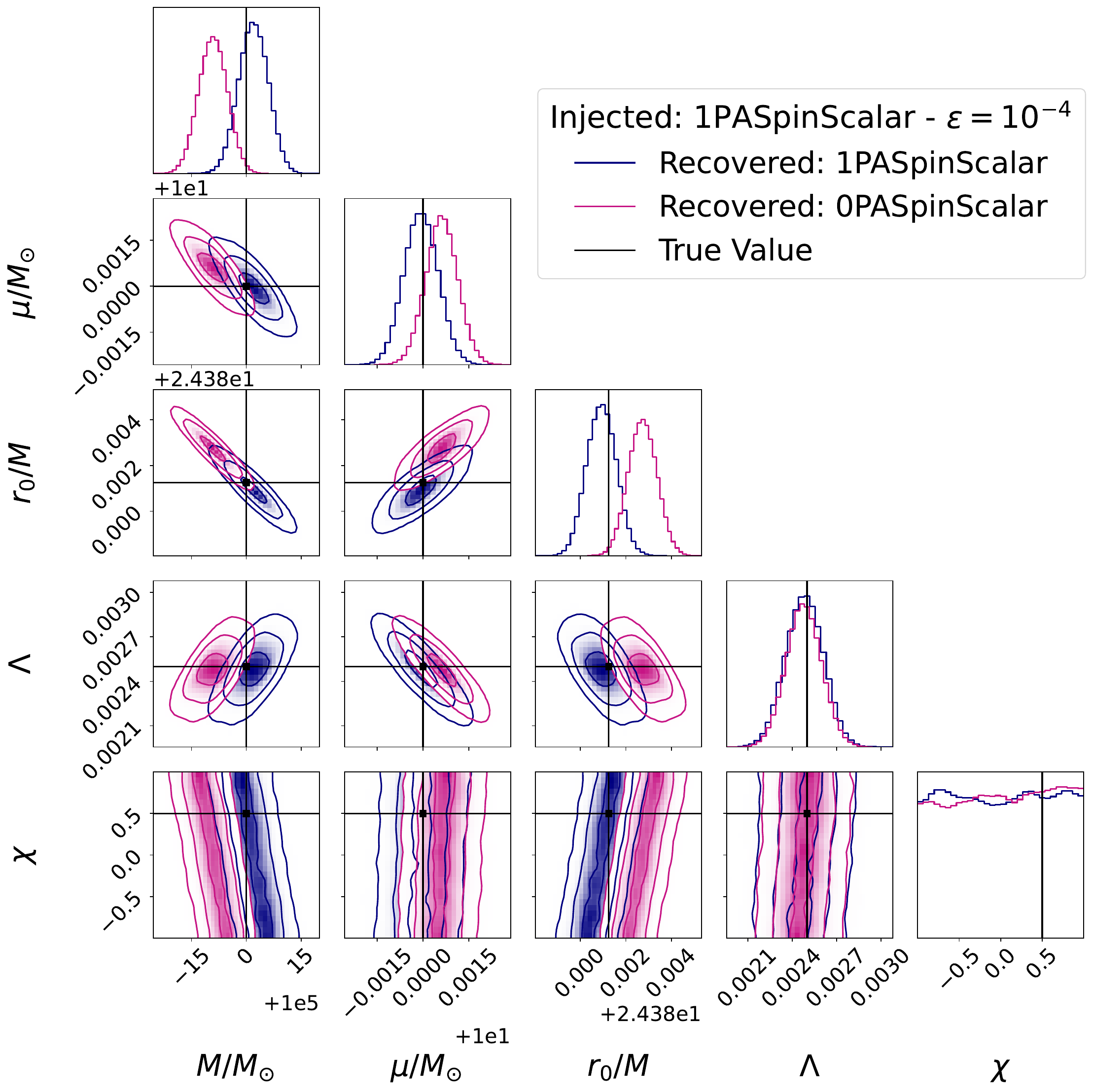}
\caption{Corner plot of the intrinsic parameters 
$(M, \mu, \hat{r}_0, \Lambda, \chi)$, obtained 
by injecting a 1PASpinScalar waveform and recovering 
with the 1PASpinScalar (blue) and 0PASpinScalar (pink) 
models. The SNR of the source is fixed to 50.
}\label{fig:4}
\end{figure}
\subsection{\label{sec:results:0PAScalarVS-1PNScalar}Case IV: 0PAScalar vs $-$1PNScalar}
As a final analysis scenario, we study the systematic 
effects that may arise when modelling scalar emission 
using a weak-field post-Newtonian (PN) approximation 
during the EMRI evolution. To this end, we replace 
the fully relativistic scalar flux $F^\Lambda_0$ with 
a PN-expanded term describing $-1$PN dipolar emission. 
We refer to this recovery template as the $-1$PNScalar 
model. In this case, the orbital evolution is governed 
by the forcing term
\begin{equation}
F_0(\hat{r}) = F^{\tn{GSF}}_0(\hat{r}) + \Lambda F^{\Lambda}_{-1\tn{PN}}(\hat{r})\,, 
\end{equation}
where\footnote{The factor $12$ is obtained by considering 
the $-1$PN scalar emission from Eq.~\eqref{eq:scalar} 
and the different normalization conventions of~\cite{Ohashi:1996uz,Castillo:2025ljw}.}
\begin{equation}
F^\Lambda_{-1\tn{PN}} = a(\hat{r}) \mathcal{F}^{\Lambda}_{-1\tn{PN}} =
a(\hat{r}) \frac{1}{12 \hat{r}^4}\,,
\end{equation}
and $a(\hat{r})$ is defined as in Section~\ref{sec:orbital_evolution}.

\begin{figure}[hbpt!]
\centering
\includegraphics[scale=0.48]{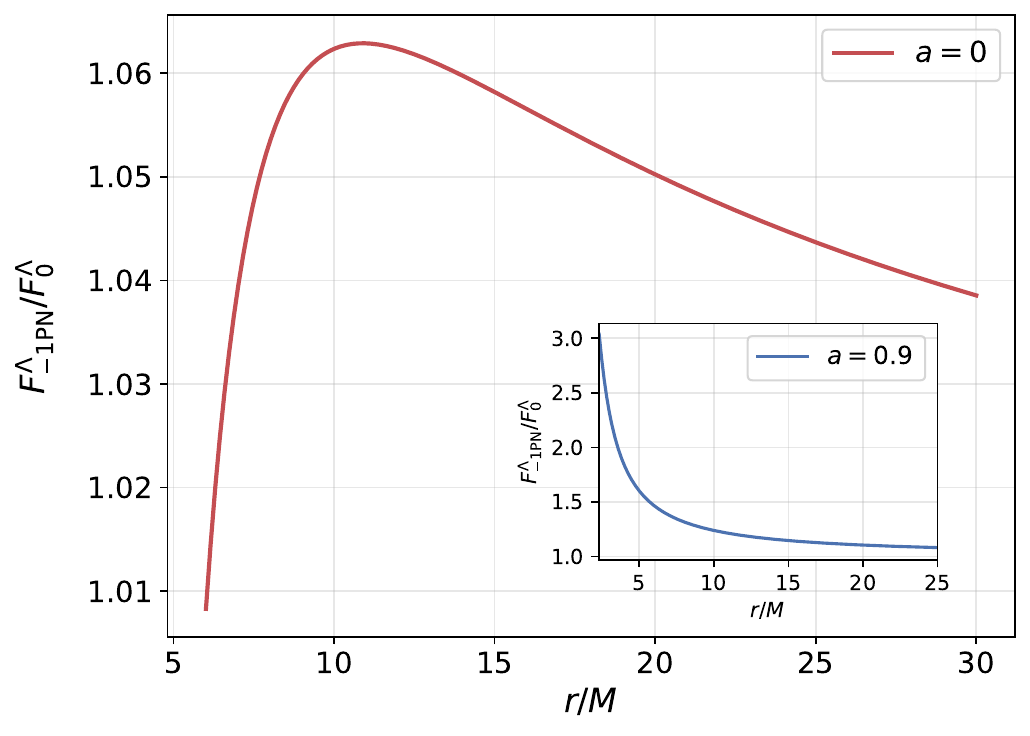}
\caption{Ratio between the $-1$PN dipolar scalar 
flux and the fully relativistic 0PA scalar flux 
as a function of the orbital radius. The inset shows the same quantity for a dimensionless primary spin  $a=0.9$.}\label{fig:5}
\end{figure}

In Fig.~\ref{fig:5} we show the 
ratio between the $-1$PN contribution and the fully 
relativistic 0PA scalar flux as a function of the 
orbital radius. The $-1$PN approximation is seen 
to overestimate the scalar flux by up to $\simeq 6\%$.
We then analyse the systematic effects arising from 
using the $-1$PNScalar approximant to recover signals 
injected with the 0PAScalar model. The corresponding 
corner plot is shown in Fig.~\ref{fig:6}, 
while the full posterior distributions are provided 
in Appendix~\ref{appendix:corner plots}, Fig.~\ref{appendix:fig:12}. The injected 
parameters are the same as those used in Figs.~\ref{fig:3} 
and~\ref{fig:4}, with $\tn{SNR} = 50$.

We find that the differences between the two models 
are negligible. In particular, we observe a bias 
of $\simeq 0.5\sigma$ in $\Lambda$ and $\mu$, and 
$\lesssim 0.3\sigma$ in all other parameters.

This result shows that modelling scalar emission with a leading-order $-1$PN approximation introduces negligible parameter biases, indicating that such simplified treatments provide an accurate description of scalar effects for the quasi-circular inspirals considered here. 
Physically, this suggests that the dominant observable imprint of the scalar sector is the presence of dipolar radiation and the associated frequency evolution, while relativistic corrections to the scalar flux contribute only subdominantly to parameter estimation.
The small impact of the relativistic corrections may be partly related to the behaviour
shown in Fig.~\ref{fig:5}: although the $-$1PN approximation systematically overestimates the fully 
relativistic scalar flux, the discrepancy remains below few percent across all the inspiral 
and decreases toward smaller orbital radii. However, this behaviour appears to be specific to inspirals
into non-spinning primaries. For Kerr black holes, the ratio between the $-$1PN and fully relativistic 
scalar fluxes  grows monotonically toward the strong-field regime, with differences larger than $50\%$ 
below $r/M\sim6$, suggesting that relativistic scalar corrections play a more important role in that case.

More generally, these findings motivate dedicated accuracy-requirement studies for scalar flux modelling, in order to quantify the level of precision needed in the scalar sector for future LISA parameter estimation analyses.

\begin{figure}[hbpt!]
\centering
\includegraphics[scale=0.22]{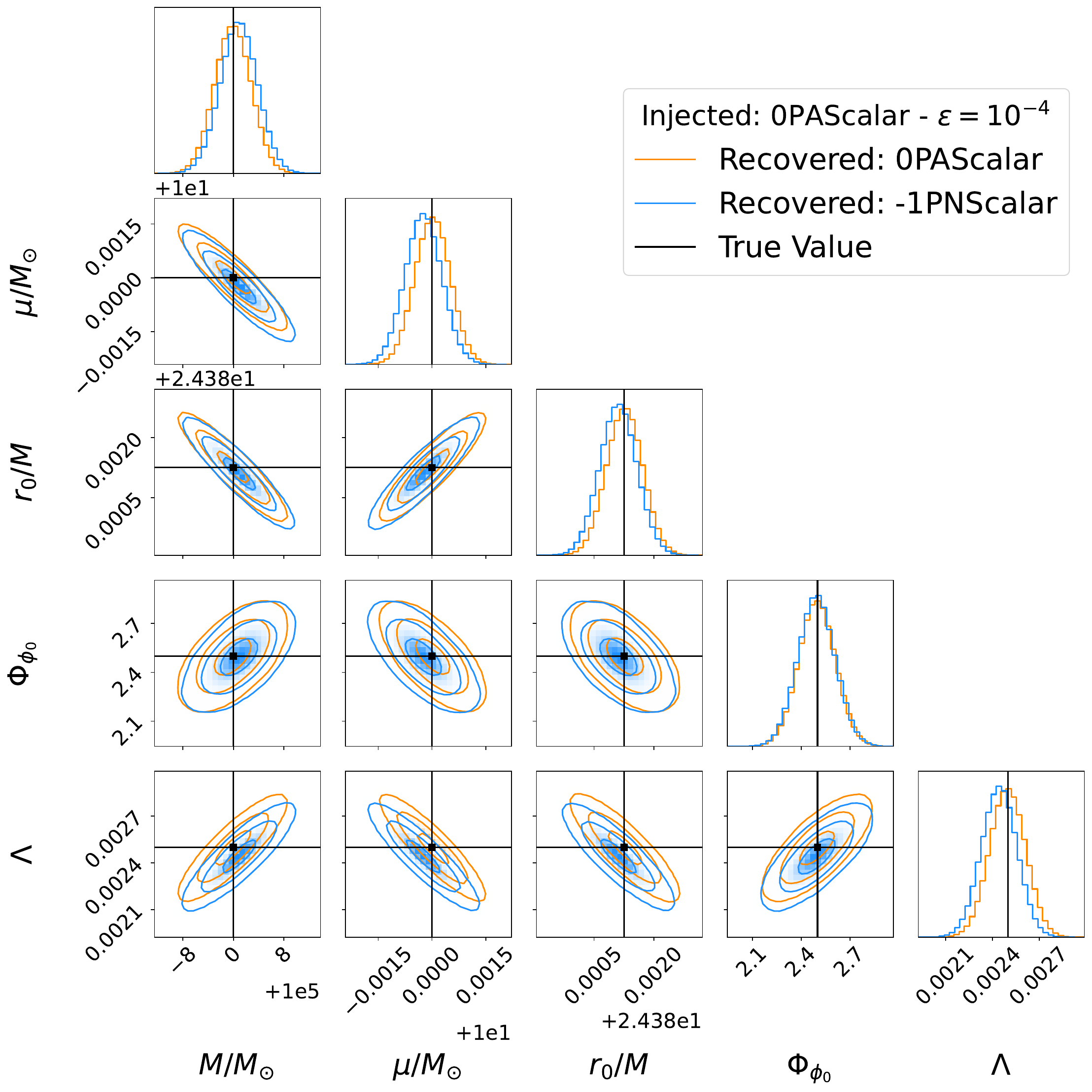}
\caption{Corner plot of the intrinsic parameters, 
showing the posterior distributions obtained by 
injecting a 0PAScalar waveform and recovering with 
the 0PAScalar (orange) and $-1$PNScalar (light blue) 
models. The injected signal has $\tn{SNR} = 50$. 
}\label{fig:6}
\end{figure}
\section{\label{sec:discussion}Discussion}
We have employed for the first time waveform models that incorporate post-adiabatic corrections computed within gravitational self-force theory to study LISA's capability of measure scalar charge for stellar-mass black holes in intermediate- and extreme-mass-ratio inspirals.
We performed a fully Bayesian analysis based on MCMC simulations, quantifying waveform systematics through a variety of injection--recovery setups and binary configurations.

We first investigated the impact of neglecting 1PA gravitational corrections while retaining adiabatic scalar and gravitational contributions in the waveform model. In this case, we found biases in the intrinsic orbital parameters, whereas the scalar charge remained largely unaffected over a broad range of mass ratios, with noticeable deviations appearing only at the largest values of $\epsilon$ or for sufficiently large scalar charges. These results suggest that, for sufficiently small mass ratios, adiabatic waveform models may already provide an adequate description for scalar-charge measurements over the one-year observation times considered here.

We then analysed the recovery of scalar signals using pure-GR templates. In this scenario, we observed large biases in both intrinsic and extrinsic parameters, together with underestimated parameter uncertainties. This result shows that neglecting scalar effects can lead to apparently precise, yet biased, parameter estimates. More broadly, it highlights the importance of incorporating non-vanilla GR effects in waveform modelling and data analysis, as accurate recovery with GR-only templates does not necessarily exclude the presence of beyond-GR physics or environmental effects.

 We also investigated the impact of including the spin of the secondary at post-adiabatic order. For the configurations considered in this work, we found that the secondary spin remains essentially unconstrained and has negligible impact on scalar-charge inference. Moreover, analyses performed in a pure-GR setup indicate that the spin of the secondary object is not measurably constrained for the mass ratio considered here, $\epsilon= 10^{-4}$, in contrast with the results reported in~\cite{Burke:2023lno}. Additional pure-GR studies presented in Appendix~\ref{appendix:secondary_spin} also show that, although the secondary spin can become partially constrained at larger mass ratios, its impact on parameter estimation is considerably smaller than that of the first post-adiabatic gravitational self-force corrections, which remain the dominant ingredient for accurate inference.

Finally, we studied the effect of modelling scalar emission using a leading-order dipolar ($-1$PN) approximation instead of fully relativistic scalar fluxes. We found negligible parameter biases in this case, indicating that such a simplified treatment accurately captures scalar effects for the quasi-circular inspirals into Schwarzschild black holes analysed here. 
This suggests that the dominant observable imprint of the scalar sector is already captured by the dipolar modification of the frequency evolution. However, it remains unclear whether this conclusion extends to inspirals into spinning primaries and generic orbits, where relativistic scalar flux corrections may become more important. Our results therefore motivate dedicated accuracy-requirement studies for scalar flux modelling aimed at quantifying the level of accuracy required in the scalar sector for future LISA data analysis.

These results should be interpreted in light of the simplifying assumptions adopted in our waveform modelling. In particular, all analyses were performed assuming observation times of $T_{\tn{obs}} = 1$ year ($1.5$ years for the pure-GR analysis with $\epsilon = 10^{-4}$). Longer observations may accumulate larger secular dephasings between waveform models and, in particular for Case I, could enhance waveform systematics and induce observable biases in the scalar charge even at smaller mass ratios.

In addition, our analysis is restricted to quasi-circular inspirals around non-rotating primary black holes, a limitation currently driven by the status of second-order self-force calculations in GR. The conclusions of this work should therefore not be directly extrapolated to generic EMRI waveforms. Rather, our results should be viewed as indicative trends and as groundwork for future studies employing more realistic waveform models. Extending the analysis to spinning primaries and generic orbits will be an important next step. Finally, incorporating post-adiabatic scalar contributions into the waveform model will be essential to fully assess the impact of scalar fields on EMRI parameter estimation.

%
\begin{acknowledgments}
 The authors acknowledge the support and computational resources provided by the Sonic high-performance computing (HPC) cluster at University College Dublin.  The authors thank Christian E.~A.~Chapman-Bird and Adam Pound for useful discussions. 
 A.S. and T.S. acknowledge partial support from the STFC Consolidated Grant no. ST/V005596/1. T.S. also acknowledges partial support from the STFC Consolidated Grants no. ST/X000672/1 and UKRI2492. O.B.~acknowledges financial support from the Grant UKRI972 awarded via the UK Space Agency. 
 A.M.~acknowledges financial support from MUR PRIN 
 Grants No.~2022-Z9X4XS and No.~2020KB33TP. A.S.~acknowledges funding from the European Union's Horizon Europe research and innovation programme under the Marie Sklodowska-Curie grant agreement no. 101199153.
 N.W.~acknowledges support from a Royal Society – Research Ireland University Research Fellowship. 
This publication has emanated from research conducted with the financial support of Research Ireland under grant number 22/RS-URF-R/3825.
\end{acknowledgments}

\appendix
\section{\label{appendix:hybrid model} Hybrid 
waveform model}
In this Appendix we describe the hybrid 1PAScalar model used for inspirals with initial radius in the range~$30<\hat{r}<60$, which we considered in Section~\ref{sec:results:1PAScalarVS0PAScalar} for $\epsilon = 4\times 10^{-4}$. 

The  orbital evolution is governed by 
\begin{equation}
\frac{d\hat{r}}{dt}  
= F^\tn{GSF}_0 + \Lambda F^{\Lambda}_0+\nu\,F^\tn{GSF}_1\,\,, \qquad 
\frac{d\Phi_\phi}{d\hat{t}} = \hat{\Omega}_\phi(\hat{r})\,.
\end{equation}
For the hybrid 1PAScalar model, the adiabatic contributions $F^\tn{GSF}_0$ and  $F^\Lambda_0$ are obtained from self-force data, numerically computed up to $\hat{r}=60$. The post-adiabatic contribution $F^\tn{GSF}_1$ is also extracted from self-force results in the range  $6.25\leq\hat{r}\leq30$, while for $\hat{r}>30$, it is obtained using 
the PN expressions for the first-order flux and energy, namely: 
\begin{equation}
    F^\tn{GSF}_{1} = - \mathcal{F}^\tn{PN}_{1} \Big[ \,
     \frac{\partial \hat{E}_0}{\partial \hat{r} }+\;\,\nu\,\frac{\partial \hat{E}^\tn{PN}_{1}}{\partial \hat{r}}\;
    \Big]^{-1}\,, 
\end{equation}
with~\cite{Honet:2025gge, Honet:2025lmk}
\begin{widetext}
\begin{align}
\mathcal{F}_1^{\rm PN}(\hat r)
=\;&
-\frac{56}{3}\,\hat r^{-6}
+\frac{37084}{315}\,\hat r^{-7}
-\frac{2332}{15}\,\pi\,\hat r^{-15/2} \nonumber
\\[6pt]
&+\hat r^{-8}\Biggl[
-\frac{134543}{1215}
+\frac{82}{15}\,\pi^2
\Biggr]+\frac{42949}{54}\,\pi\,\hat r^{-17/2} \nonumber
\\[6pt]
&+\hat r^{-9}\Biggl[
-\frac{1452202403629}{229209750}
+\frac{1327296}{1225}\,\gamma_E
-\frac{267127}{720}\,\pi^2
+\frac{15329984}{11025}\,\ln 2
+\frac{37908}{49}\,\ln 3
+\frac{663648}{1225}\,\ln\!\Bigl(\frac{1}{\hat r}\Bigr)
\Biggr] \nonumber
\\[6pt]
&+\hat r^{-19/2}\Biggl[
\frac{2062241}{3465}\,\pi
+\frac{328}{15}\,\pi^3
\Biggr]\,, 
\end{align}
and
\begin{equation}
\hat{E}^\tn{PN}_{1}(\hat{r}) =
\frac{1}{24\hat{r}^2}
- \frac{19}{16\hat{r}^3}
+ \frac{5}{1152}\left(-6889 + 246\pi^2\right)\frac{1}{\hat{r}^4}
- \frac{1}{46080\hat{r}^5}
\left(
-494684 + 1376256\,\gamma_E + 135555\,\pi^2
+ 688128 \ln\!\left(\frac{16}{\hat{r}}\right)
\right)\,.
\end{equation}
\end{widetext}
Here $\gamma_E = 0.5772...$ denotes the Euler--Mascheroni constant. 
\section{\label{appendix:corner plots} Corner plots of the full posterior distributions}
\begin{figure*}[hbpt]
\centering
\includegraphics[scale=0.26]{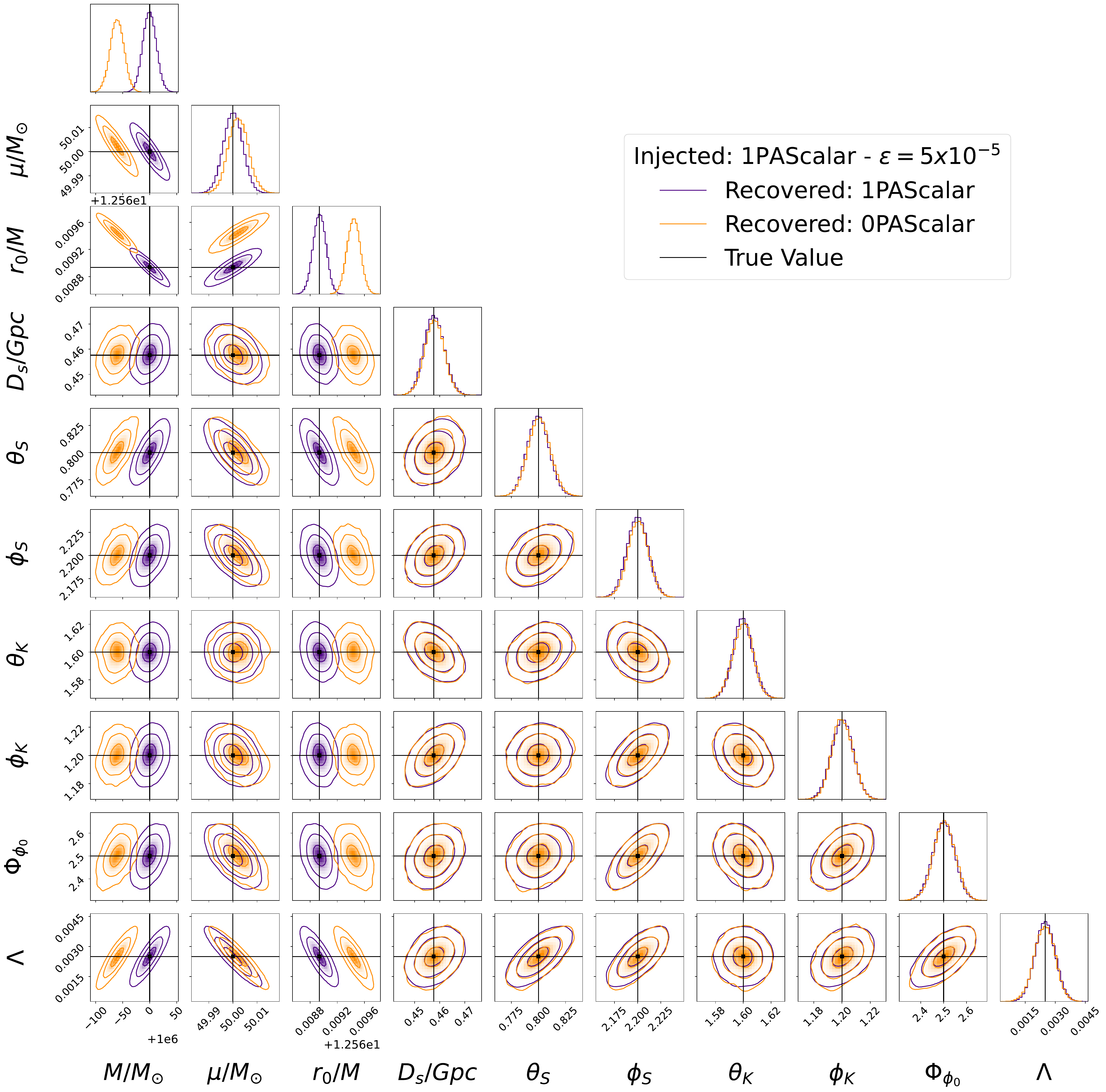}
\caption{Corner plot of the ten binary parameters for $\epsilon = 5\times10^{-5}$, ontained by injecting a 1PAScalar waveform model and recovering with the same model (purple) and the 0PAScalar template (orange). 
The time of observation is fixed to $1$ year and the SNR is fixed to $200$.}\label{appendix:fig:7}
\end{figure*}

\begin{figure*}[hbpt]
\centering
\includegraphics[scale=0.26]{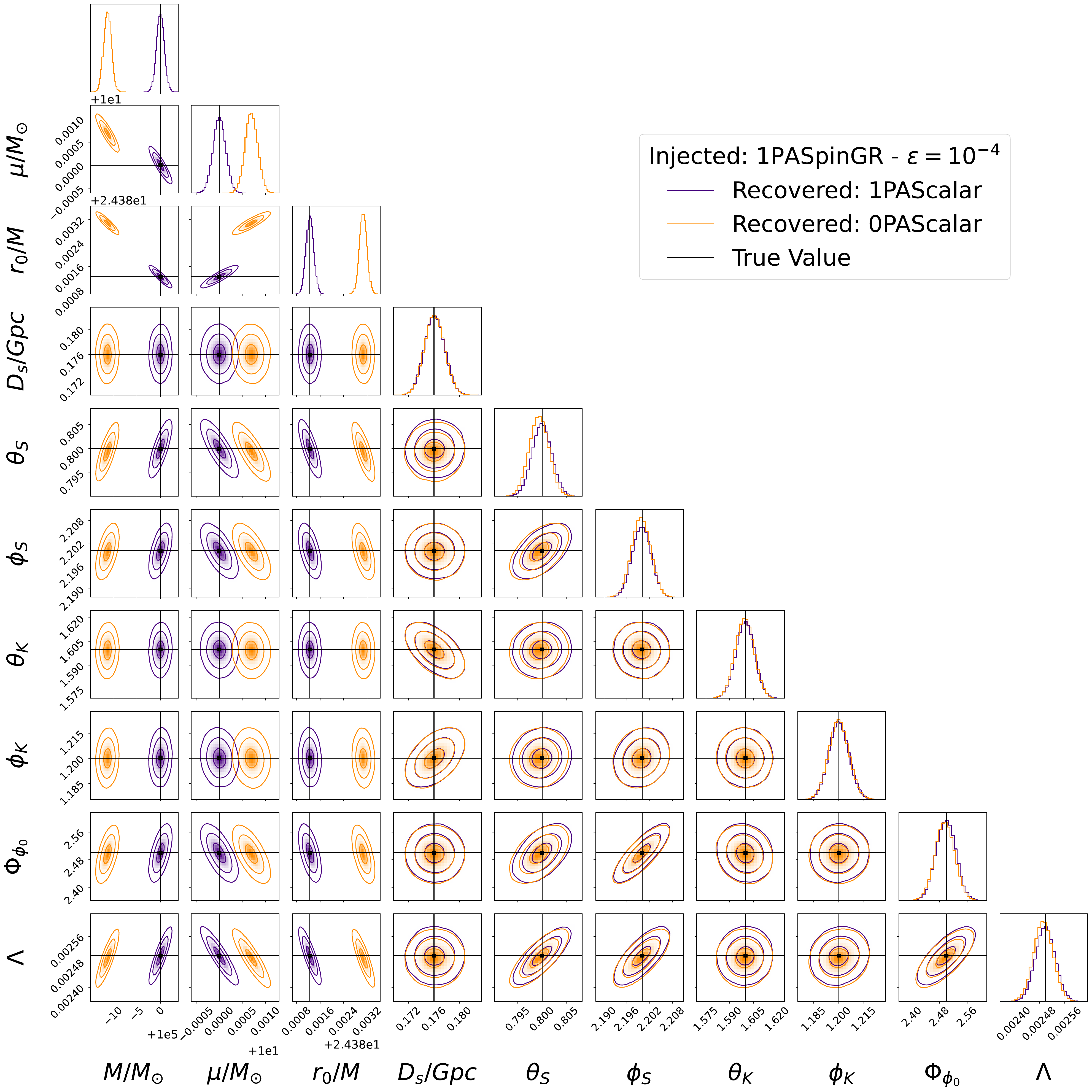}
\caption{Same as Fig.~\ref{appendix:fig:7} but for $\epsilon = 10^{-4}$.
The SNR is fixed to $200$.}\label{appendix:fig:8}
\end{figure*}

In this Appendix we present the full corner plots of the systematic analyses. Figs.~\ref{appendix:fig:7}, \ref{appendix:fig:8} and \ref{appendix:fig:10} are associated with Section~\ref{sec:results:1PAScalarVS0PAScalar} for $\epsilon = 5\times10^{-5},10^{-4}$ and $4\times10^{-4}$, respectively. Fig.~\ref{appendix:fig:11} shows the corner plot for the systematic analysis including the secondary spin in Section~\ref{sec:results:1PASpinScalarVS0PASpinScalar}, and Fig.~\ref{appendix:fig:12} shows the plot for the study regarding the dipolar emission in Section~\ref{sec:results:0PAScalarVS-1PNScalar}.
\section{\label{appendix:secondary_spin} Reassessment of post-adiabatic effects in pure-GR parameter estimation}
In this appendix, we present a set of pure-GR analyses aimed at reassessing the measurability of the secondary spin and comparing our results with those reported in~\cite{Burke:2023lno}. Specifically, we consider: (i) a repetition of the analysis shown in Fig.~4 of~\cite{Burke:2023lno}, corresponding to a mass ratio $\epsilon=10^{-4}$; (ii) a repetition of the analysis shown in Fig.~5 of~\cite{Burke:2023lno}, corresponding to $\epsilon=10^{-3}$; and (iii) an additional study investigating the impact of neglecting the secondary spin in the recovery waveform. In all three cases, the injected angular parameters are $(\theta_S, \phi_S, \theta_K, \phi_K, \Phi_{\phi_0}) = (\pi/3, \pi/4, 2.0, 5.0, 1.5)$.

First, we revisit the configuration considered in Fig.~4 of~\cite{Burke:2023lno}. We inject and recover signals using the same waveform model, denoted 1PASpinGR, which includes both the first post-adiabatic gravitational self-force and first post-adiabatic secondary spin contributions in the trajectory evolution, namely the forcing terms $F^\tn{GSF}_{0}$, $F^\tn{GSF}_{1}$, and $F^{\chi}_1$ appearing in Eq.~\eqref{eq:fluxes}.

We consider the same initial configuration as in Fig.~4 of~\cite{Burke:2023lno}, with $\hat{r}_0=15.7905$, $M=10^6 M_\odot$, $\mu=10^2 M_\odot$, $T_\tn{obs} = 1.5$ years and luminosity distance $D_S = 2\,\tn{Gpc}$, corresponding to a signal with $\tn{SNR}\simeq 80$. The posterior distributions are shown in the corner plot in Fig.~\ref{appendix:fig:13}.
As discussed in Section~\ref{sec:results:1PASpinScalarVS0PASpinScalar}, we observe that the secondary spin remains unconstrained. 

We then consider the case with mass-ratio $\epsilon=10^{-3}$, for a comparison with Fig.~5 in~\cite{Burke:2023lno}, where we injected the 1PASpinGR model and we performed the recovery by using both the same model and the 0PASpinGR template. Here, the initial orbital radius is $\hat{r}_0\simeq16.809$, component masses are $M=5\times10^6 M_\odot$ and $\mu=5\times10^3 M_\odot$, the observation time is $T_\tn{obs} = 1$ year and $\tn{SNR}\simeq 340$, corresponding to a signal with luminosity distance $D_S = 0.4\,\tn{Gpc}$. The posterior distributions are shown in Fig.~\ref{appendix:fig:14}.

 In this case, when recovering with the same model used for the injection, we find a posterior for the secondary spin that is qualitatively similar to that reported in~\cite{Burke:2023lno}: the distribution exhibits some structure rather than being completely unconstrained. A more notable difference emerges when using the 0PASpinGR template. In this case, the resulting posterior closely resembles that obtained in~\cite{Burke:2023lno} when including the 1PA self-force corrections through the 3PN approximation. This is in contrast with the findings of~\cite{Burke:2023lno}, where the corresponding analysis including only the 0PA self-force contribution and the 1PA secondary spin effects produced an essentially flat posterior on the secondary spin.

 Finally, we investigate the impact of neglecting the secondary spin in the recovery model. To this end, we inject a signal generated with the 1PASpinGR waveform and recover it using three templates: 1PASpinGR, 1PAGR, and 0PAGR. The initial orbital radius is set to $\hat{r}_0\simeq 14.477$, the primary mass is $M=10^6 M_\odot$ and the secondary mass is $\mu = 100 M_\odot$. The time of observation is fixed to $T_\tn{obs}=1$ year, so that the final radius of the 1PASpinGR trajectory is $\simeq 6.35$. The signal-to-noise ratio is fixed to $\tn{SNR}=70$, giving a luminosity distance of $D_S\simeq1.3$ Gpc. 

When recovering with the 1PAGR template, the posterior distributions remain broadly consistent with those obtained using the full 1PASpinGR model, although small offsets are visible in the inferred values of $M$ and $\hat{r}_0$. These offsets are nevertheless much smaller than those observed when using the 0PAGR template, for which significant biases affect the intrinsic parameters $M$, $\hat{r}_0$ and $\mu$. 

The configuration considered here is similar to that shown in the second panel of Fig.~2 of~\cite{Burke:2023lno}, although we adopt a shorter observation time ($T_\tn{obs} = 1$ year instead of $T_\tn{obs}=1.5$ years). Despite this difference, we find qualitatively similar behaviour: neglecting the secondary spin contribution leads only to modest shifts in the recovered intrinsic parameters, whereas omitting the 1PA self-force corrections produces substantially larger biases.

Overall, these results suggest that the 1PA self-force corrections play a much more important role in mitigating parameter biases than the inclusion of the secondary spin contribution for the configuration considered here.

\begin{figure*}[hbpt]
\centering
\includegraphics[scale=0.26]{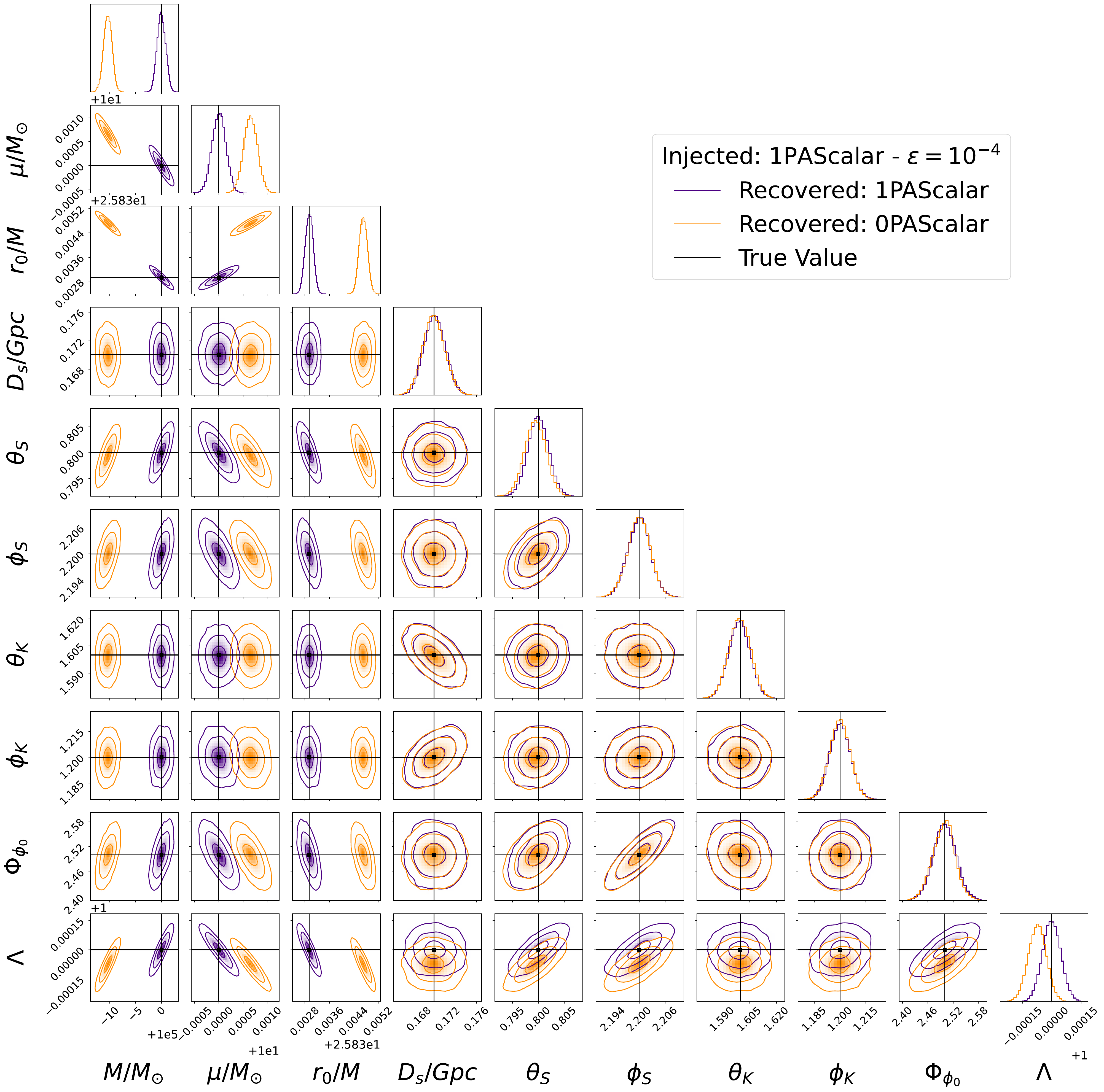}
\caption{Same as Fig.~\ref{appendix:fig:8} but for scalar charge $d = 1$. 
The SNR is fixed to $200$.}\label{appendix:fig:9}
\end{figure*}
\begin{figure*}[hbpt]
\centering
\includegraphics[scale=0.26]{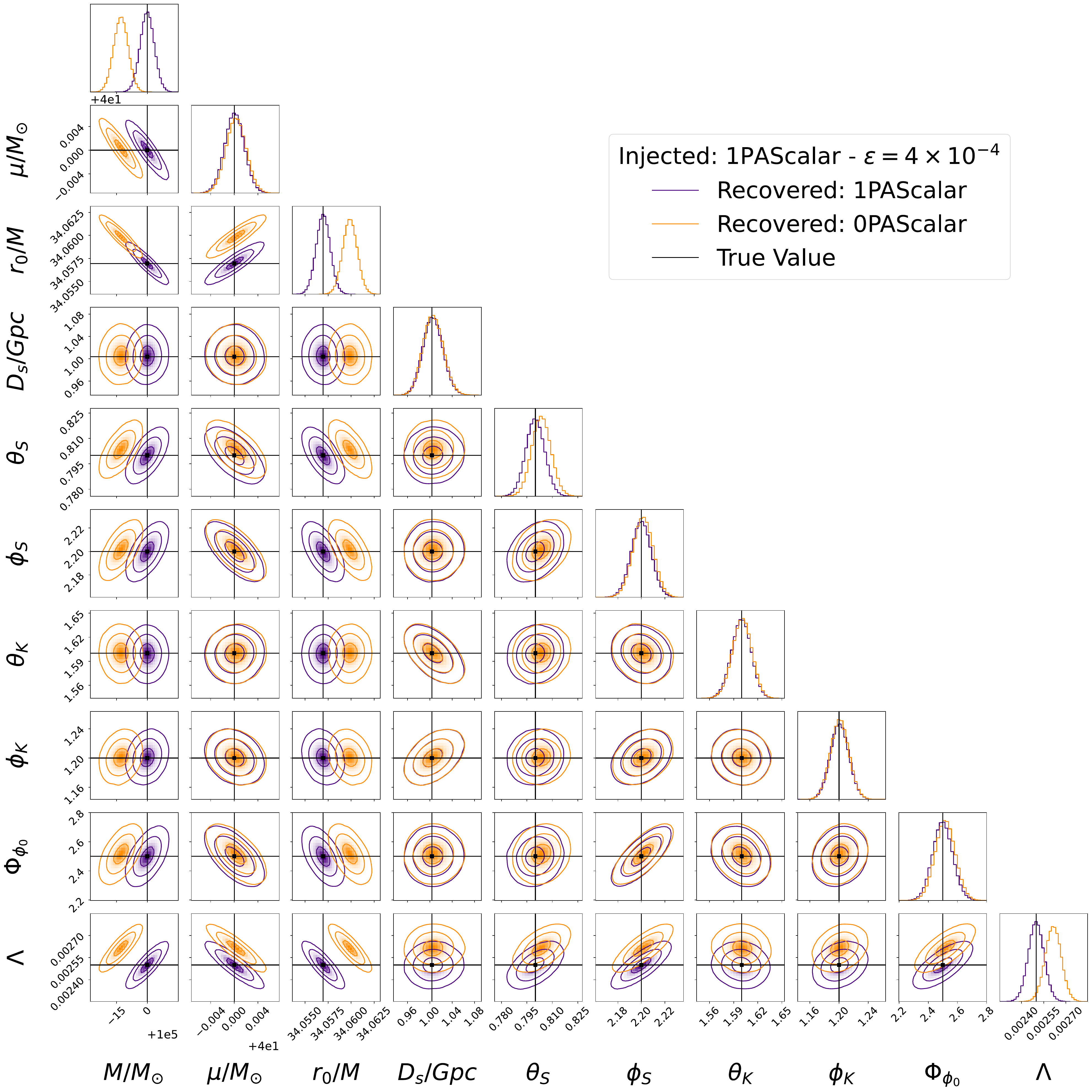}
\caption{Same as Fig.~\ref{appendix:fig:7} but for $\epsilon = 4\times 10^{-4}$ and SNR=100. 
}\label{appendix:fig:10}
\end{figure*}
\begin{figure*}[hbpt]
\centering
\includegraphics[scale=0.26]{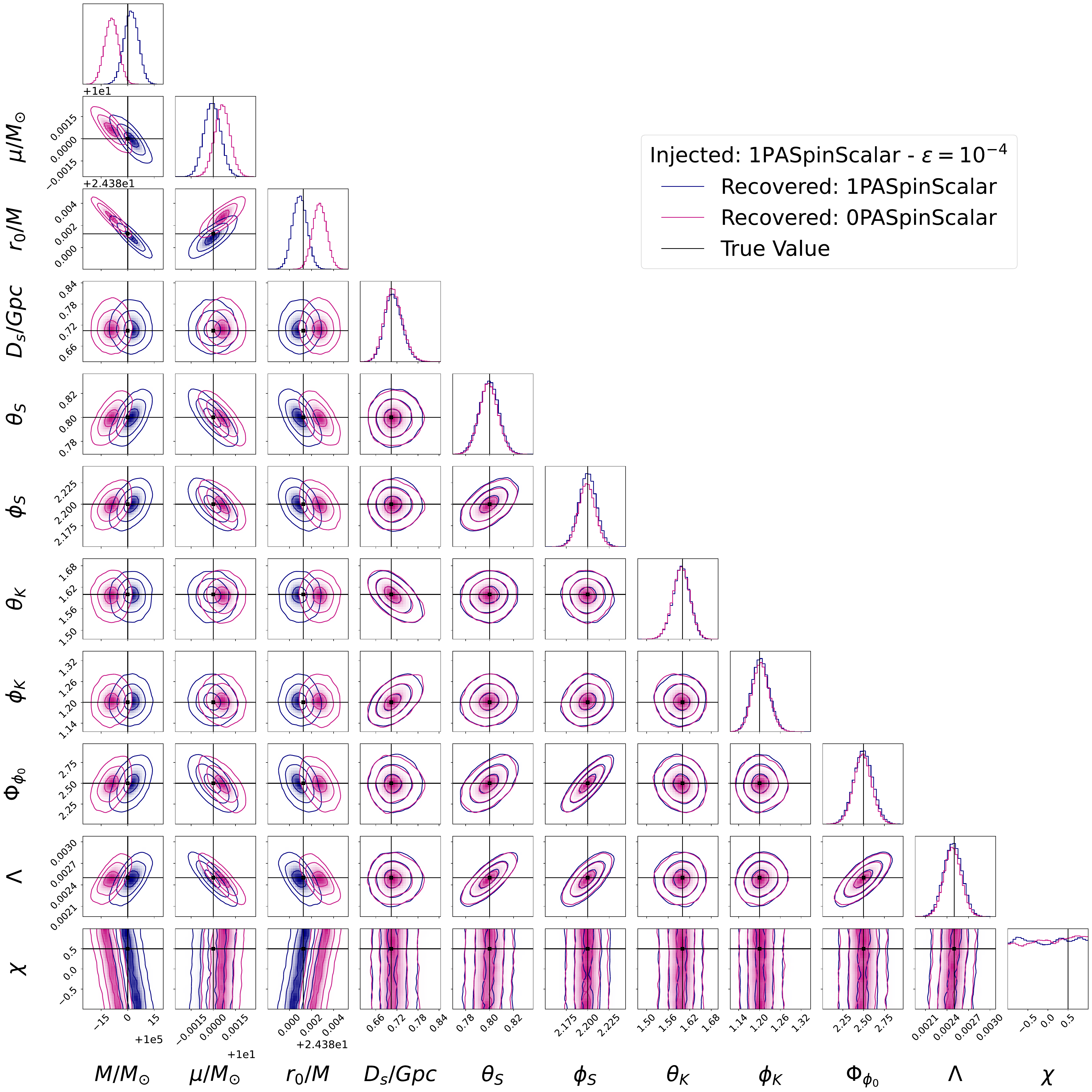}
\caption{Corner plot of the binary parameters obtained by injecting a 1PASpinScalar model and recovering with the same model (indigo) and the 0PASpinScalar template (pink). 
The observation time is fixed to $T_\tn{obs}=1$ year, and the signal-to-noise ratio to SNR=50.  }\label{appendix:fig:11}
\end{figure*}
\begin{figure*}[hbpt]
\centering
\includegraphics[scale=0.26]{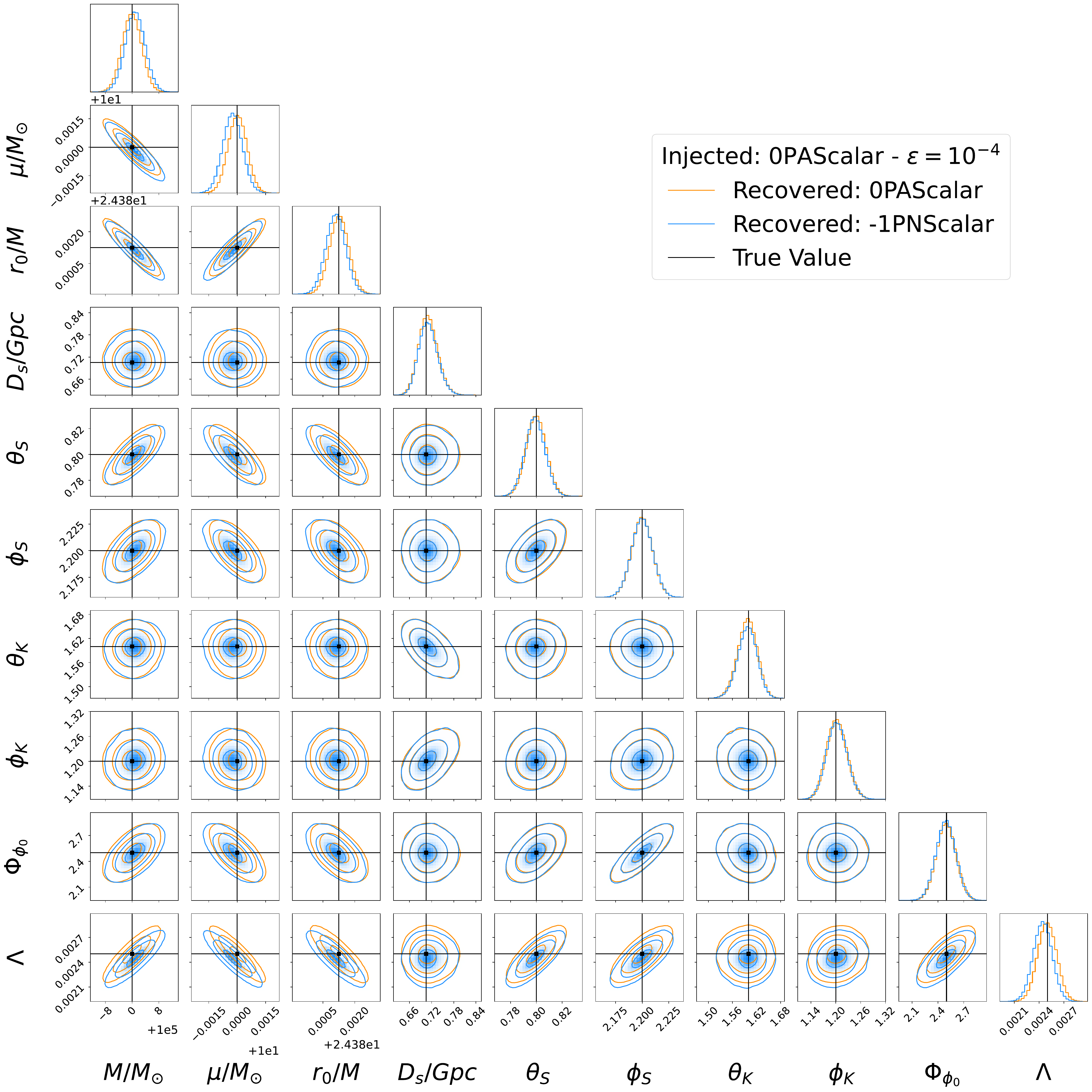}
\caption{Corner plots showing the posterior distributions obtained by injecting a 0PAScalar waveform template and recovering with the same model (light blue) and the $-1$PNScalar template. The observation time is fixed to $T_\tn{obs}=1$ year and SNR=50.  }\label{appendix:fig:12}
\end{figure*}
\begin{figure*}[hbpt]
\centering
\includegraphics[scale=0.26]{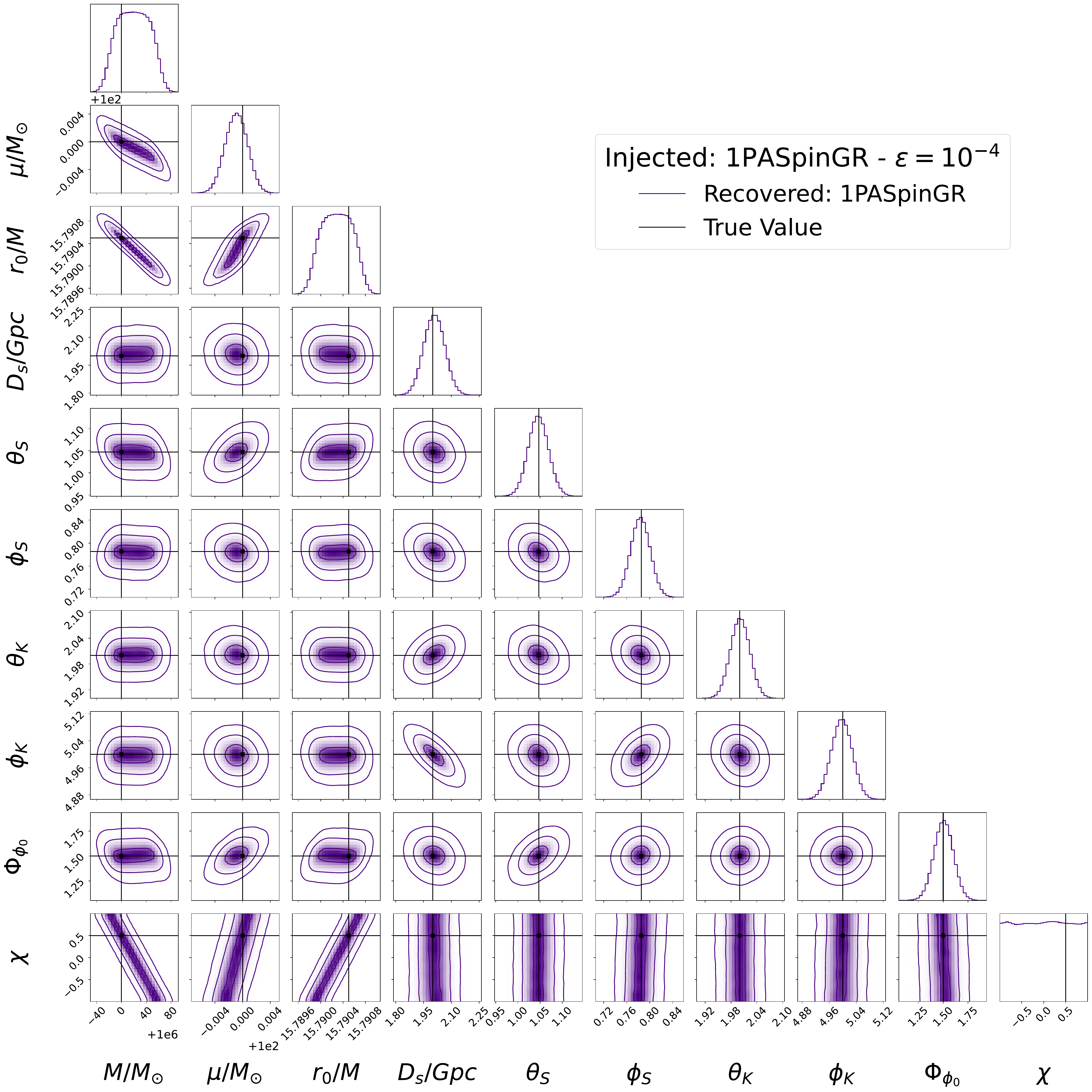}
\caption{Corner plots showing the posterior distributions obtained by injecting a 1PASpinGR waveform template and recovering with the same model (purple). The intial orbital radius is $\hat{r}_0 = 15.7905$. The observation time is fixed to $T_\tn{obs}=1.5$ years and the injected luminosity distance is $D_S= 2~\tn{Gpc}$, corresponding to a signal with $\tn{SNR}\simeq80$.  }\label{appendix:fig:13}
\end{figure*}
\begin{figure*}[hbpt]
\centering
\includegraphics[scale=0.26]{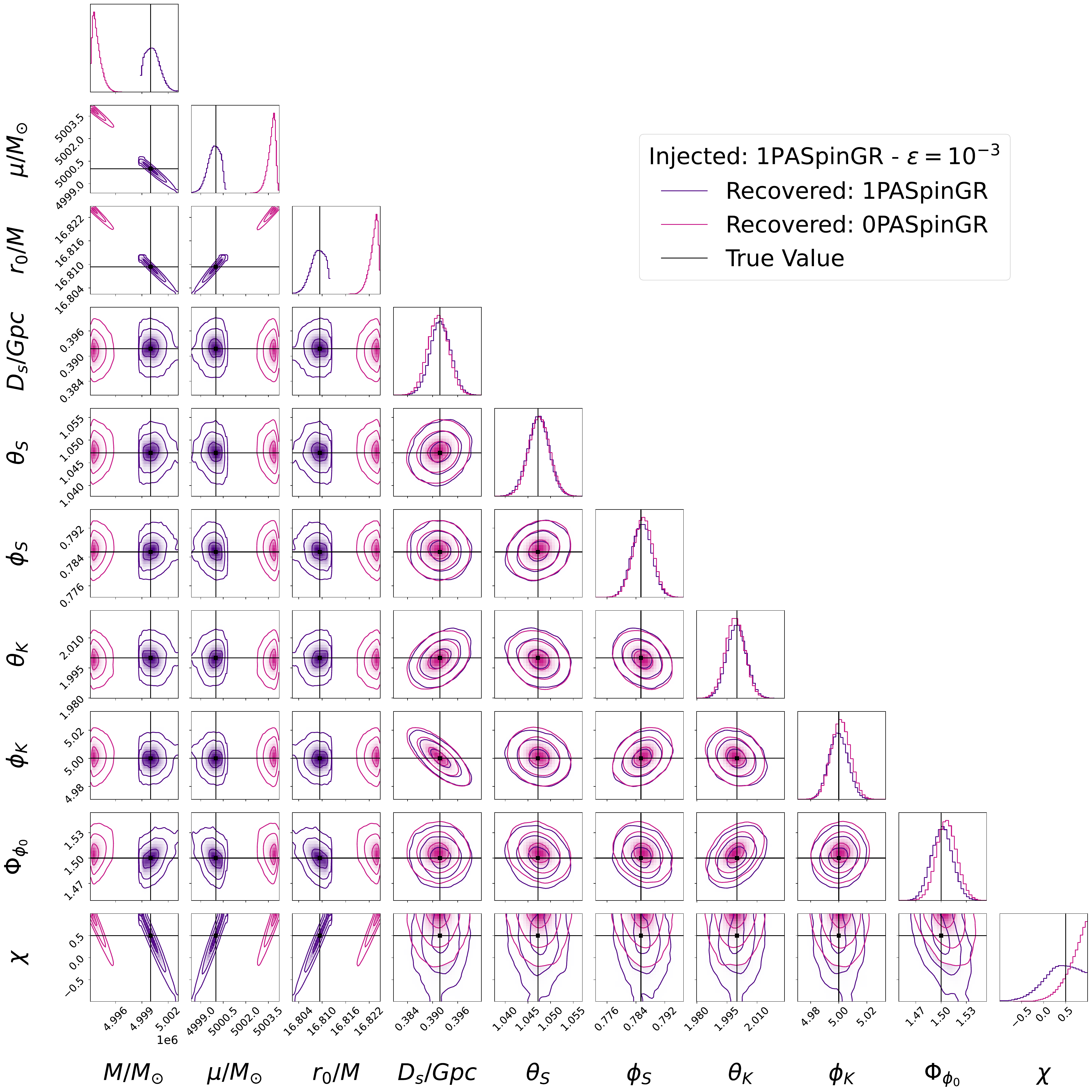}
\caption{Corner plots showing the posterior distributions obtained by injecting a 1PASpinGR waveform template and recovering with the same model (purple) and the 0PASpinGR template (pink). The intial orbital radius is $\hat{r}_0 \simeq 16.805$. The observation time is fixed to $T_\tn{obs}=1$ year and the $\tn{SNR}$ is fixed to $340$, corresponding to a signal with luminosity distance $D_S\simeq0.4~\tn{Gpc}$.  }\label{appendix:fig:14}
\end{figure*}
\begin{figure*}[hbpt]
\centering
\includegraphics[scale=0.26]{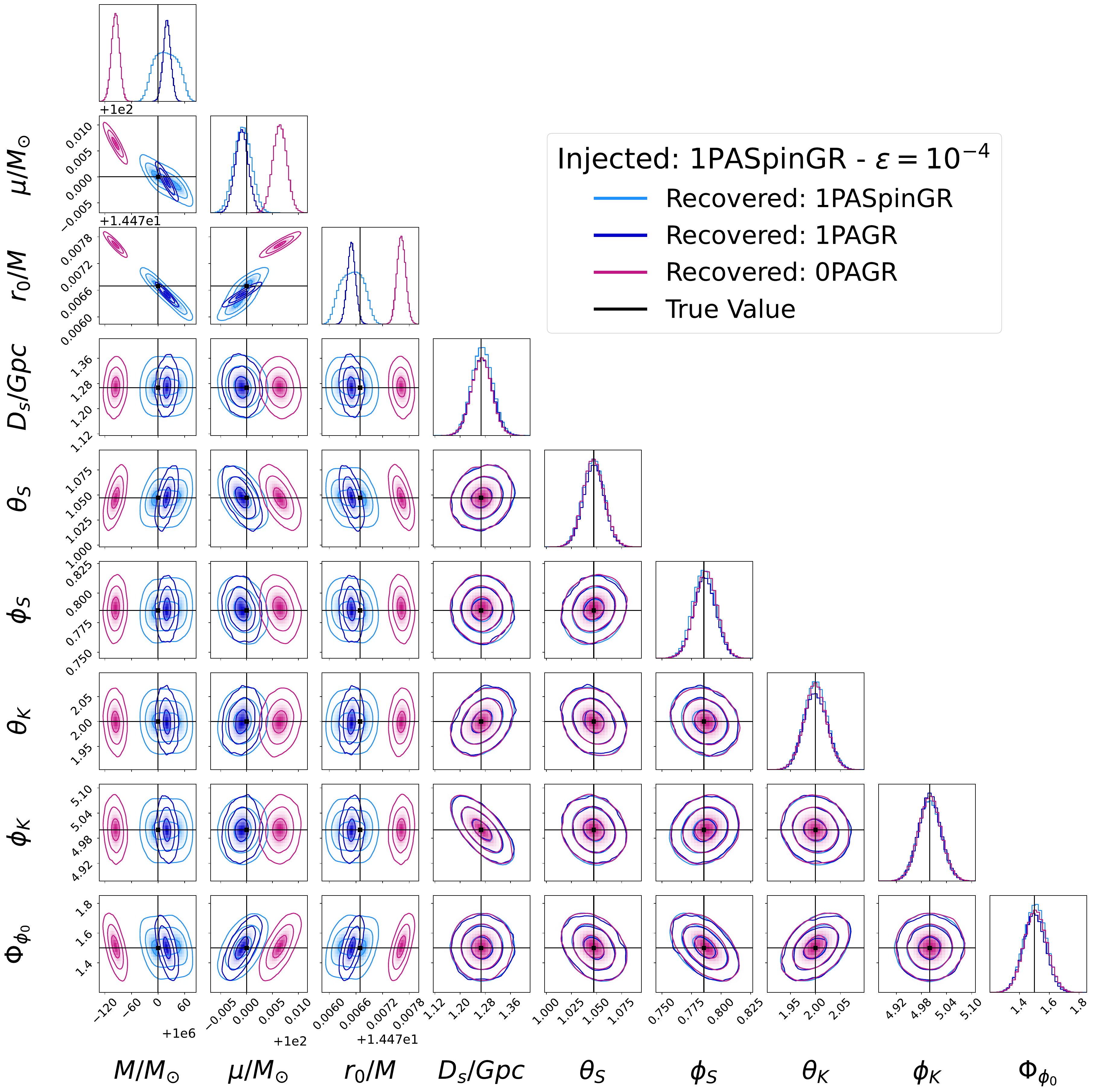}
\caption{Corner plots showing the posterior distributions obtained by injecting a 1PASpinGR waveform template and recovering with the same model (light blue), the 1PAGR (blue) and the 0PAGR (pink) templates. The intial orbital radius is $\hat{r}_0 \simeq 14.477$. The observation time is fixed to $1$ year and the $\tn{SNR}$ is fixed to $70$, corresponding to a signal with luminosity distance $D_S\simeq1.3~\tn{Gpc}$.  }\label{appendix:fig:15}
\end{figure*}
\clearpage
\bibliography{bib.bib}
\bibliographystyle{apsrev4-1}
\end{document}